%% file: FCAL_JINST.tex
\documentclass[cits]{JINST}
\usepackage{graphicx}
\usepackage{epsfig}
\usepackage{subfig}
\usepackage{psfrag}

\newcommand{\subfigure}{\subfloat}
\newcommand{\Subref}[1]{\protect\subref{#1}}

\def\etal{et~al.}%
\newcommand{\mc}[1]{\mathcal{#1}}
\title{Forward Instrumentation for ILC Detectors}

\author{Halina Abramowicz$^a$, Angel Abusleme$^b$, Konstantin Afanaciev$^c$, 
Jonathan Aguilar$^d$, Prasoon Ambalathankandy$^d$,
 Philip Bambade$^e$, Matthias Bergholz$^{f,1}$,
Ivanka Bozovic-Jelisavcic$^g$, 
Elena Castro$^f$, Georgy Chelkov$^h$, Cornelia Coca$^i$, 
Witold Daniluk$^j$, 
 Angelo Dragone$^k$, Laurentiu Dumitru$^i$,
Konrad Elsener$^l$,
Igor Emeliantchik$^c$, Tomasz Fiutowski$^d$, Mikhail Gostkin$^h$,
Christian Grah$^{f,2}$,
Grzegorz Grzelak$^{j,3}$, Gunther Haller$^k$,
Hans Henschel$^f$, 
Alexandr Ignatenko$^{c,4}$, 
Marek Idzik$^d$,
Kazutoshi Ito$^m$, Tatjana Jovin$^g$,
Eryk Kielar$^j$,
Jerzy Kotula$^j$,
Zinovi Krumstein$^h$, Szymon Kulis$^d$,
Wolfgang Lange$^f$, Wolfgang Lohmann$^{f,1}$\thanks{Corresponding
author.},
Aharon Levy$^a$, Arkadiusz Moszczynski$^j$,
Uriel Nauenberg$^n$, Olga Novgorodova$^{f,1}$, Marin Ohlerich$^{f,1}$, 
Marius Orlandea$^i$,
Gleb Oleinik$^n$, 
Krzysztof Oliwa$^j$,
Alexander Olshevski$^h$, Mila Pandurovic$^g$, Bogdan Pawlik$^j$, 
Dominik Przyborowski$^d$, Yutaro Sato$^m$,
Iftach Sadeh$^a$, Andre Sailer$^l$, Ringo Schmidt$^{f,1}$,
Bruce Schumm$^o$,
Sergey Schuwalow$^f$, Ivan Smiljanic$^g$, Krzysztof Swientek$^d$, Yosuke Takubo$^m$,
Eliza Teodorescu$^i$,
Wojciech Wierba$^j$, Hitoshi Yamamoto$^m$, Leszek Zawiejski$^j$ 
and Jinlong Zhang$^p$ \\
\llap{$^a$}Tel Aviv University, Tel Aviv, Israel\\
\llap{$^b$}Stanford University, Stanford, USA\\
\llap{$^c$}NCPHEP, Minsk, Belarus \\
\llap{$^d$}AGH University of Science \& Technology, Cracow, Poland\\
\llap{$^e$}Laboratoire de l Accelerateur Lineaire, Orsay, France\\
\llap{$^f$}DESY, Zeuthen, Germany\\
\llap{$^g$}Vinca Institute of Nuclear Sciences, University of Belgrade, Serbia\\
\llap{$^h$}JINR, Dubna, Russia \\
\llap{$^i$}IFIN-HH, Bucharest, Romania\\
\llap{$^j$}INP  PAN, Cracow, Poland \\
\llap{$^k$}SLAC, Menlo Park, USA\\
\llap{$^l$}CERN, Geneva, Switzerland\\
\llap{$^m$}Tohoku University, Sendai, Japan\\
\llap{$^n$}University of Colorado, Boulder, USA\\
\llap{$^o$}UC California, Santa Cruz, USA\\
\llap{$^p$}ANL, Argonne, USA\\
\llap{$^1$}also at Brandenburg University of Technology, Cottbus, Germany\\
\llap{$^2$}now at BTO Consulting AG, Berlin, Germany\\
\llap{$^3$}also at University of Warsaw, Poland\\
\llap{$^4$}now at DESY, Hamburg, Germany\\
E-mail: \email{Wolfgang.Lohmann@desy.de}\\}

\abstract{Two special calorimeters are foreseen for the instrumentation of
the very forward region of
the ILC detector,  
a luminometer  designed to measure the rate of
low angle Bhabha scattering events with a precision better than 10$^{-3}$ and
a low polar angle calorimeter, adjacent to the beam-pipe. The latter will be hit by a large amount 
of beamstrahlung remnants. The amount and shape of these
depositions will allow a fast luminosity estimate
and the determination of beam parameters. The sensors
of this calorimeter must be radiation hard.
Both devices will improve the hermeticity of the detector in the search for new particles.
Finely segmented and very compact calorimeters will match the
requirements. Due to the high occupancy 
fast front-end electronics is needed.
The design of the calorimeters developed and optimised with Monte Carlo
simulations is presented. Sensors and readout electronics ASICs have been designed
and prototypes are available. Results on the performance of these major components are  
summarised.}


\begin{document}

\section{Introduction and challenges}

A high energy e$^+$e$^-$ linear collider 
is considered to be the future research facility complementary to the LHC collider.
Whereas LHC has a higher potential for discoveries, an e$^+$e$^-$ collider will allow
precision measurements to explore in detail the mechanism of 
electroweak symmetry breaking and the properties of the physics beyond the Standard Model, should it be found at the LHC.
Two concepts of an e$^+$e$^-$ linear collider are presently considered, the ILC~\cite{ILC_pub}
and CLIC~\cite{clic_info}.
For the ILC, with superconducting cavities, an engineering design report will 
be issued in 2012. The centre-of-mass energy will be
500 GeV, with the possibility of an upgrade to 1 TeV. CLIC is based on 
conventional cavities. A conceptional design report is foreseen in 2011.
CLIC will allow to collide electrons and positrons up to energies of 3 TeV.  

An R\&D program is ongoing to develop the technologies for detectors 
for precision measurements in this new energy domain. 
Letters of Intent have been submitted for detectors at the ILC in 2009. Two detectors, 
the ILD~\cite{ILD_pub} and the
SiD~\cite{SiD_pub}, are reviewed and validated. 
In both detectors
two specialised calorimeters are foreseen in the very forward region, LumiCal for the precise measurement
of the luminosity and BeamCal for a fast estimate of the 
luminosity and for the control of beam parameters~\cite{ieee1}. 
Both will also improve the hermeticity of the detector. 
To support beam-tuning an additional pair-monitor will be positioned just 
in front of BeamCal.

With LumiCal 
the luminosity will be measured using
Bhabha 
scattering, ${\rm{e}}^+{\rm{e}}^- \rightarrow {\rm{e}}^+{\rm{e}}^-(\gamma)$, as a gauge process.
To match the physics benchmarks,
an 
accuracy of better than
10$^{-3}$ is needed at a centre-of-mass energy of 500 GeV~\cite{ILD_pub}. 
For the GigaZ option, where the ILC will be operated for precision
measurements 
at centre-of-mass energies around the Z boson, an accuracy
of 10$^{-4}$ would be required~\cite{klaus}. To reach these accuracies,
a precision device is needed,
with particularly challenging  
requirements on the mechanics 
and position control.

BeamCal is positioned
just outside the beam-pipe.
At ILC energies we have to tackle here a new phenomenon -- the
beamstrahlung.
When electron and positron bunches
collide, 
the particles are accelerated in the magnetic field of the bunches
towards the bunch centre. This so called pinch effect enhances the 
luminosity. However, 
electrons and positrons
may radiate photons. A fraction of these 
photons converts in the Coulomb field of
the bunch particles creating low energy ${\rm{e}}^+{\rm{e}}^-$ pairs. 
 A large amount of these pairs 
will deposit their energy after each bunch crossing
in BeamCal. These depositions, useful for a 
bunch-by-bunch luminosity estimate and the determination of 
beam parameters~\cite{grah1}, 
will lead, however, to a 
radiation dose of about one MGy per year in the sensors
at lower polar angles.
Hence radiation hard sensors are needed to instrument 
BeamCal. 
BeamCal is supplemented by a pair monitor, consisting of a layer 
of pixel sensors positioned 
just in front of it to measure the density of 
beamstrahlung pairs and
give additional information for the beam parameter determination.

All detectors in the very forward region have to tackle relatively high 
occupancy, requiring special front-end electronics.

A small Moli\`{e}re radius is of importance for both calorimeters. 
It ensures high energy
electron veto capability for BeamCal even
at small polar angles. This is essential to suppress
background in searches for new particles
for which the signature 
consists of large missing energy and momentum.
In LumiCal the precise reconstruction of electron, positron and photon 
showers in Bhabha events
is facilitated.
Both calorimeters also shield the inner tracking detectors from back-scattered particles
induced by beamstrahlung pairs hitting the downstream  beam-pipe and magnets.
\begin{figure}[b]
\centerline{\includegraphics[width=0.6\columnwidth]{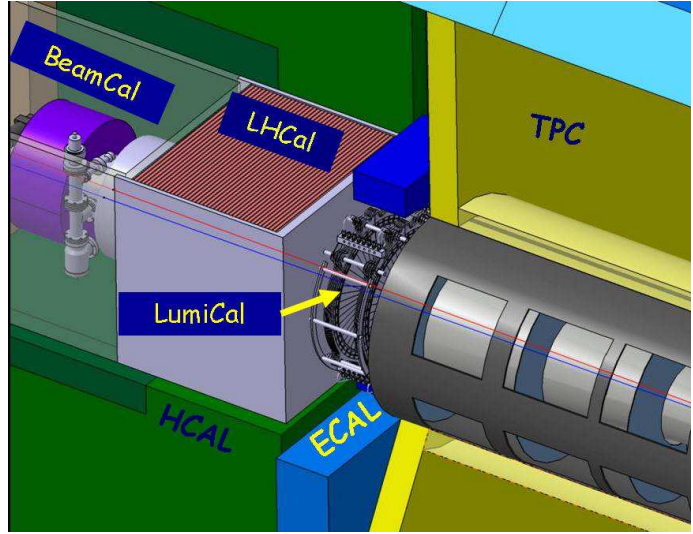}}
\caption{The very forward region of the ILD detector. LumiCal, BeamCal and LHCal are carried by 
the support tube for the final focusing quadrupole and the beam-pipe. 
LHCal extends the coverage of the hadron calorimeter 
to the polar angle range of LumiCal.
TPC denotes the central track chamber, ECAL the electromagnetic and 
HCAL the hadron calorimeter.\label{fig:Forward_structure}}
\end{figure}

\input{design_jinst.tex}

\input{sys_eff.tex}

\input{sensors_jinst.tex}

\input{asic_lumi_jinst.tex}

\input{BeamCal_ASIC.tex}

\input{pair_moi_asic.tex}

\section{Summary}
A design for the instrumentation of the very forward region of a detector at the International 
Linear collider is presented. Two calorimeter are planned, LumiCal to measure precisely the 
luminosity and BeamCal, supplemented by a pair monitor, for a fast luminosity estimate and beam tuning.
Both calorimeters extend the coverage of the detector 
to small polar angles. Parameters relevant for the physics program 
have been estimated by Monte Carlo simulations and found to match
the requirements for the chosen
geometry.
Prototypes of the major components such as sensors, front-end ASICs and ADC ASICs are developed,
produced and tested. Their measured performance fulfils the specifications 
derived from the
Monte Carlo simulations.
The results presented here demonstrate that  the sensors and the ASICs are ready to be integrated  into 
a fully functional prototype detector and to perform, as the next step, tests of 
fully assembled sensor plane prototypes.

\section{Acknowledgments}

This work is supported by the Commission of the European Communities
under the 6$^{th}$ Framework Program "Structuring the European
Research Area", contract number RII3-026126.
Tsukuba University is supported in part by the Creative 
Scientific Research Grant No. 18GS0202 of the Japan 
Society for Promotion of Science.
The AGH-UST is supported by the Polish Ministry of Science and Higher
Education under contract Nr. 372/6.PRUE/2007/7.
The INP PAN is supported by
    the Polish Ministry of Science
    and Higher Education
    under contract Nr. 141/6.PR UE/2007/7.
IFIN-HH is supported by the Romanian Ministry of Education, Research and Innovation 
through the Authority CNCSIS under contract IDEI-253/2007.
The VINCA group is benefiting from the project "Physics and Detector R$\&$D in 
HEP Experiments" supported by the Ministry of Science of the Republic of Serbia.
J. Aguilar, P. Ambalathankandy and O. Novgorodova 
are supported by the 7th Framework Programme
"Marie Curie ITN", grant agreement number 214560.

%


\end{document}

%% file: design_jinst.tex
\section{Design of the very forward region} 
 
A sketch of the very forward region of the ILD detector~\cite{ILD_pub}
is 
shown in 
Figure~\ref{fig:Forward_structure}. 
LumiCal and BeamCal are designed as cylindrical 
sensor-tungsten sandwich
electromagnetic calorimeters.
Both consist of 30 absorber disks of 3.5 mm thickness, each corresponding to
one radiation length, interspersed with sensor 
layers.
Each sensor layer is segmented radially and azimuthally 
into pads.
 Front-end ASICs are positioned at the outer radius
of the calorimeters.
LumiCal is positioned in a circular hole of the
end-cap electromagnetic calorimeter ECAL.
BeamCal is placed just in front of the final focus quadrupole.
BeamCal covers polar angles between 5 and 40 mrad and LumiCal between
31 and 77 mrad. 

Colliding beams enter the interaction point, IP, with a crossing angle of 
14 mrad. Both calorimeters are centred around the outgoing beam. 
In the design of BeamCal a hole for the incoming beam-pipe is foreseen.

\subsection{LumiCal simulation studies} 

The differential cross section of 
Bhabha scattering,
$\frac{d\sigma_{\mathrm{B}}}{d\theta}$,
can be calculated precisely 
from theory~\cite{new_Bhabha}.
In leading order it reads,
\begin{equation}{
\frac{d\sigma_{\mathrm{B}}}{d\theta} =
\frac{2\pi \alpha^{2}_{\rm{em}}}{s} \frac{\sin \theta}{\sin ^{4}(\theta / 2)} \approx 
\frac{32\pi \alpha^{2}_{\rm{em}}}{s} \frac{1}{\theta^{3}} \; ,
}\label{bhabhaXs2EQ} \end{equation}
where $\theta$ is the polar angle of the scattered electron with respect to the beam.
The approximation holds at small $\theta$.

\begin{figure}[htpb]
\centering
\subfigure[]{
\label{fig:bhabhaXSFIG1}
\includegraphics[trim=0mm 15mm 0mm 25mm,clip,width=.45\textwidth]{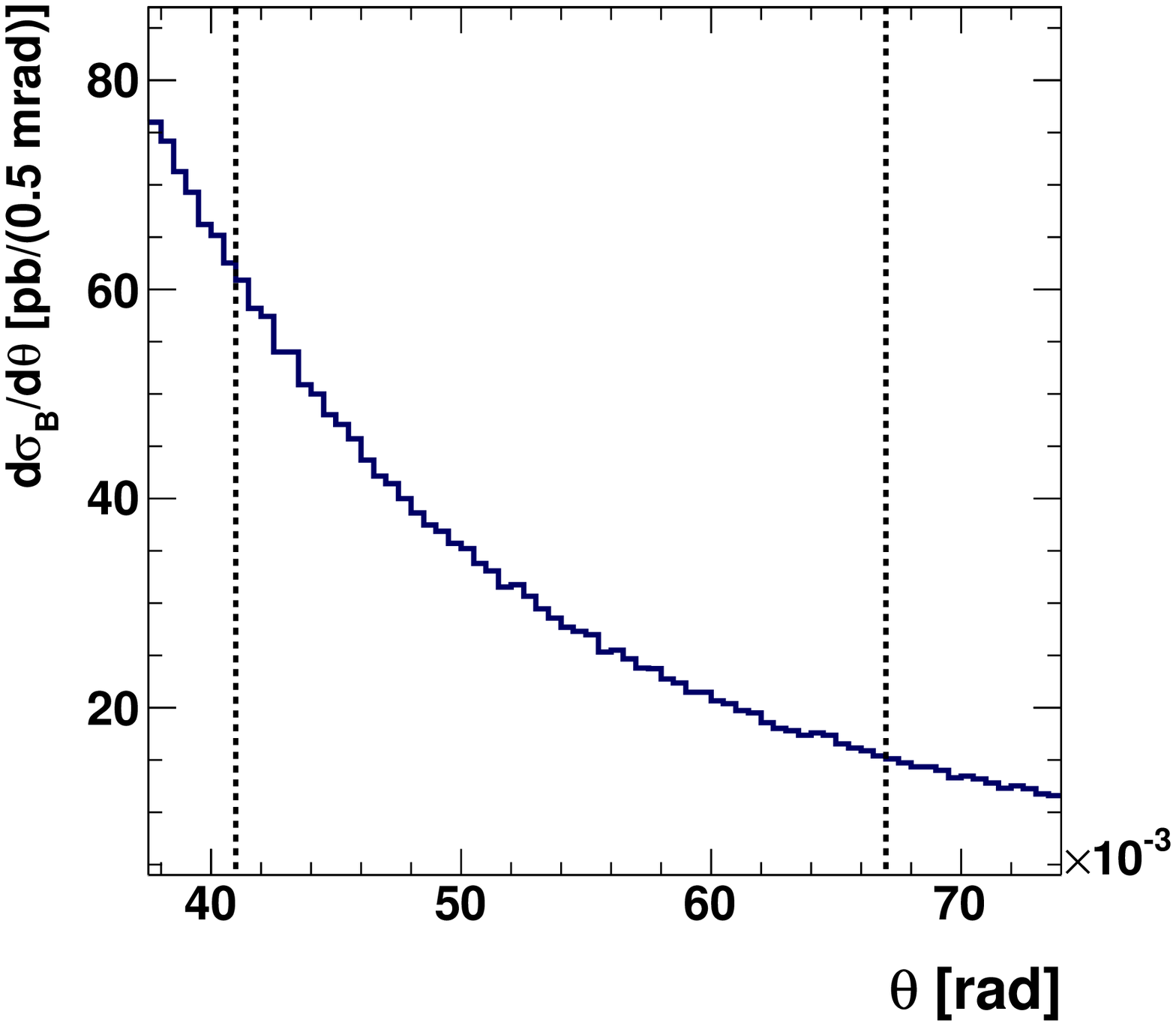}}
\subfigure[]{
\label{fig:bhabhaXSFIG2}
\includegraphics[trim=0mm 15mm 0mm 25mm,clip,width=.45\textwidth]{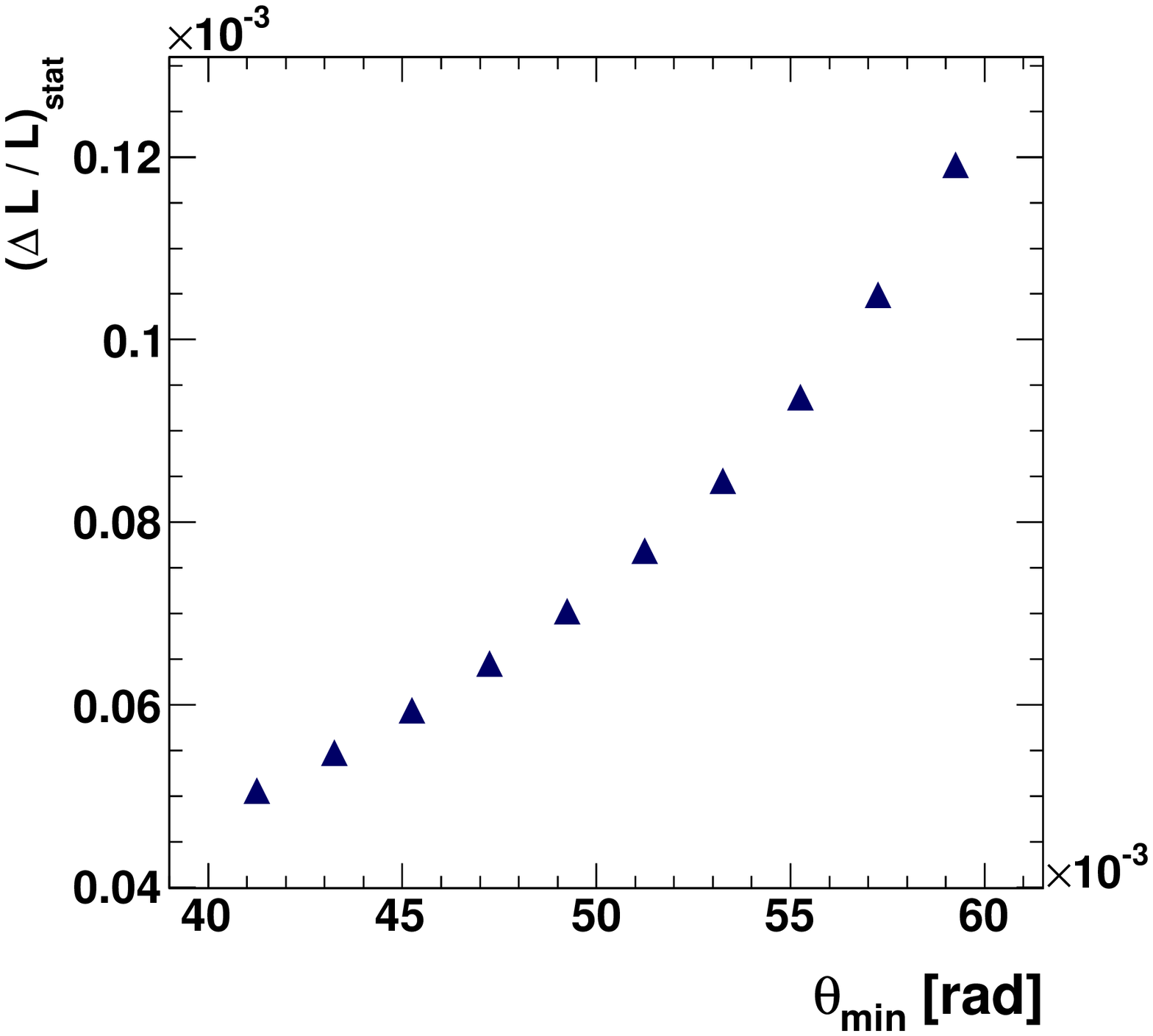}}
\caption{\Subref{fig:bhabhaXSFIG1} Dependence of $d\sigma_{\mathrm{B}}/d\theta$, the differential Bhabha 
cross-section, on the polar angle, $\theta$, at $\sqrt{s}=500$~GeV. 
The dashed lines mark the fiducial volume of LumiCal, 
$41<\theta<67$~mrad, which is defined in eqn.~(2.5) later in this chapter.
\Subref{fig:bhabhaXSFIG2} Dependence of the statistical uncertainty in counting the number of 
Bhabha events, $(\Delta \sf{L} / \sf{L})_{stat}$, on the minimal polar angle of the fiducial volume, 
$\theta_{\rm{min}}$, 
while
the upper limit is kept at 67 mrad. An integrated luminosity of 500 fb$^{-1}$ is assumed.
}
\label{fig:bhabhaXSFIG}
\end{figure}

For a given rate of Bhabha events, N$_{\mathrm{B}}$, determined in a certain 
$\theta$-range,
the luminosity, $\sf{L}$, is obtained as
\begin{equation}{
{\sf{L}} = \frac{{\rm{N}}_{\mathrm{B}}}{\sigma_{\mathrm{B}}},
}\label{lumiDefEQ} \end{equation}
where $\sigma_{\mathrm{B}}$ is the integral of the differential cross section, eqn.~(\ref{bhabhaXs2EQ}),
over the considered $\theta$ range. 
Because of the steep $\theta$ dependence of the cross section, 
as illustrated in   
Figure~\ref{fig:bhabhaXSFIG1}, 
the most critical quantity to control when counting 
Bhabha events 
is the inner acceptance radius 
of the calorimeter, defined as the lower cut in the polar 
angle,
$\theta_{\rm{min}}$. 
Hence a very precise  $\theta$ measurement is needed.
Furthermore, the $\theta$-range must be chosen such that
the
number of Bhabha events measured
provides the required relative
statistical uncertainty of 10$^{-3}$. 
By choosing the lower bound of the polar angle between 
40 and 60 mrad the latter requirement can be easily reached
as illustrated in Figure~\ref{fig:bhabhaXSFIG2}. 
Here a Bhabha event sample generated with the BHWIDE generator~\cite{BHWIDE}
was used. 
The generated sample corresponds to an integrated luminosity of 500 fb$^{-1}$, as 
expected in one year 
of 
running 
the collider at nominal luminosity.  

Electromagnetic showers are simulated in LumiCal using the GEANT4~\cite{g4} based package 
Mokka~\cite{mokka}. Sensors consist of 300 $\mu$m thick silicon sectors covering an azimuthal 
angle of 30$^\circ$.
The depositions in each sensor pad are recorded, and a reconstruction of the 
shower is performed.
The position of an electromagnetic shower in LumiCal is reconstructed by performing 
a weighted average over the 
energy deposits in
 individual pads. 
The weight, $\mc{W}_{\rm{i}}$, of a given detector pad i is 
determined by logarithmic weighting~\cite{bib17},  for which $\mc{W}_{\rm{i}} = \mathrm{max} \{~ 0 ~,~ \mc{C} + \mathrm{ln} ({\rm{E}}_{\rm{i}} 
/ {\rm{E}}_{\rm{tot}} ~) \}$. 
Here ${\rm{E}}_{\rm{i}}$ refers to the individual pad energy, 
${\rm{E}}_{\rm{tot}}$ is the total energy in all pads, and $\mc{C}$ is a constant. 
In this way,
only pads which contain a sufficient fraction of the shower energy 
contribute to the reconstruction. 
The polar angle resolution, $\sigma_{\theta}$, and a polar angle measurement bias, 
$\Delta \theta$, are defined as the Gaussian width and 
the central value of the difference 
between the reconstructed and the generated polar angles. 
There is an optimal value 
for $\mc{C}$, for which $\sigma_{\theta}$ is minimal~\cite{bib20,bib23}. 

Non-zero values of $\Delta \theta$ are due to the non-linear
signal sharing on finite size pads  with gaps between them. 
The bias and the resolution in the polar angle measurement depend on the polar angle pad size. 
The bias causes a shift in the 
luminosity measurement, since events may migrate into or out 
of the fiducial volume. This shift reads as
\begin{equation}
\left( \frac{\Delta\sf{L}}{\sf{L}} \right)_{\rm{rec}} \approx 2 \frac{\Delta \theta}{\theta_{\rm{min}}} \;.
\label{luminosityRelativeErrRec2EQ} 
\end{equation}
\begin{figure}[htpb]
\begin{center}
\subfigure[]{
\label{lumiBiasThetaRecFIG2}
\includegraphics[clip,width=.45\textwidth]{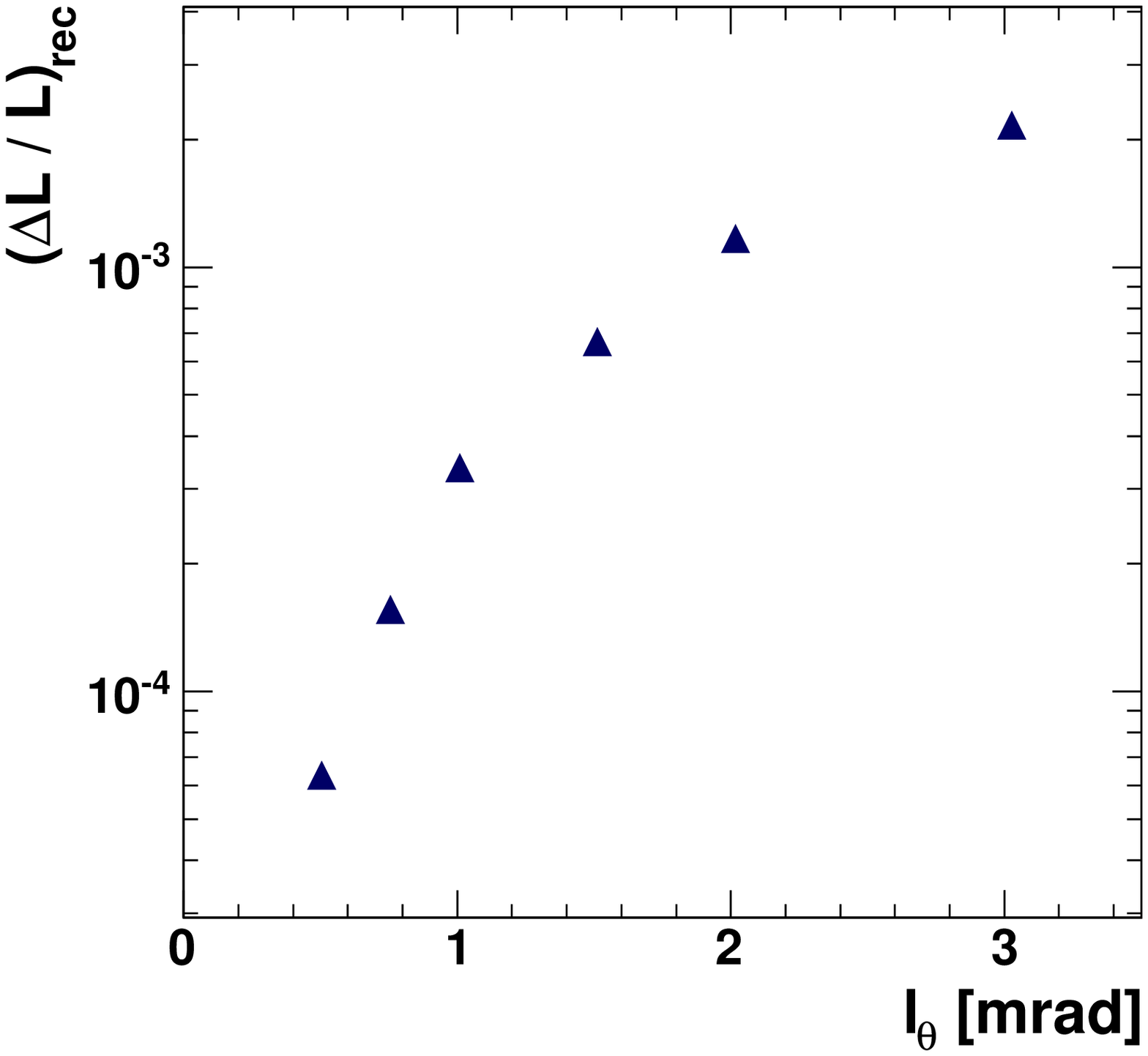}} 
\subfigure[]{
\label{engyResThetaMinMaxFIG1}
\includegraphics[clip,width=.5\textwidth,height=.43\textwidth]{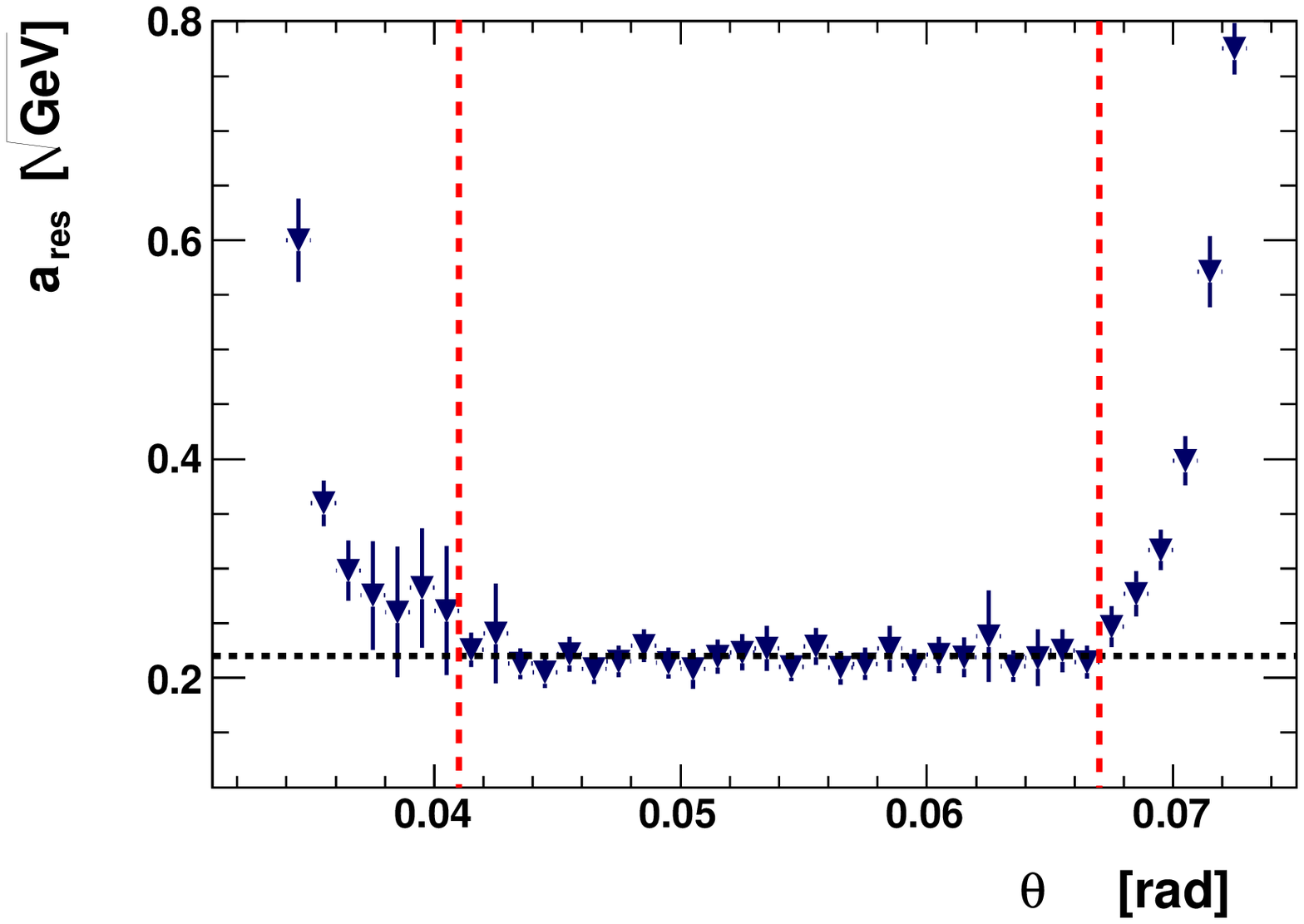}}
\caption{\Subref{lumiBiasThetaRecFIG2} 
Dependence of $\left( \Delta\sf{L} / \sf{L} \right)_{\rm{rec}}$, 
as defined in eqn.~(2.3), on 
the polar angle pad size, l$_{\theta}$. 
\Subref{engyResThetaMinMaxFIG1} 
The energy resolution, a$_{\rm{res}}$, for 250~GeV 
electrons as a function of the polar angle, $\theta$,
covering the polar angle range of the LumiCal.}
\end{center}
\end{figure}
Figure~\ref{lumiBiasThetaRecFIG2} 
shows the relative shift in the luminosity as a function 
of the polar 
angular pad size, l$_{\theta}$, using the optimal value of $\mc{C}$. 
For l$_{\theta} < 2$~mrad 
the shift in the luminosity measurement is smaller than $10^{-3}$. 
As the baseline for the design we have chosen
l$_{\theta} = 0.8$~mrad, 
which corresponds to 64~radial divisions of the sensor. 
For this segmentation
the polar angle
resolution and bias amount 
to $\sigma_{\theta} = (2.2\, \pm\, 0.01) \times 10^{-2}$
and $\Delta \theta = (3.2\, \pm\, 0.1) \times 10^{-3}$~mrad, respectively.
The relative shift in 
the luminosity is $\left( \Delta\sf{L} / \sf{L} \right)_{\rm{rec}} = 1.6 \times 10^{-4}$. 

The polar angle bias needs careful understanding
in
test-beam measurements 
with
sensors finally chosen for the calorimeter. 
Once its value is known,  a correction can be applied to the luminosity 
measurement.
The 
uncertainty of the luminosity measurement is then given by the uncertainty of the 
measured bias which may be smaller than the shift itself.
The value of $1.6 \times 10^{-4}$ can therefore be considered as an upper 
bound on the relative luminosity bias.

With
30 radiation lengths of tungsten as
absorber, high energy 
electrons and photons deposit almost all of their energy 
in the detector. 
The relative energy 
resolution, $\sigma_{\rm{E}} /{\rm{E}}$, is parametrised as
\begin{equation}{
\frac{\sigma_{\rm{E}}}{\rm{E}} = \frac{{\rm{a}}_{\rm{res}}}{\sqrt{{\rm{E}}_{\rm{beam}}~\mathrm{(GeV)}}},
}\label{engyResEQ} \end{equation}
\noindent where ${\rm{E}}$ and $\sigma_{\rm{E}}$ are, respectively, 
the central value and the standard deviation of 
the distribution of the energy deposited in the sensors for a beam of electrons with 
energy $\rm{E}_{\rm{beam}}$. The parameter ${\rm{a}}_{\rm{res}}$ is 
usually quoted as the energy resolution, a convention which will be followed here.

Figure~\ref{engyResThetaMinMaxFIG1} shows the energy resolution 
as a function of the polar angle $\theta$ for electron showers with energy 
250~GeV. 
The energy resolution parameter approaches minimal constant values between  
$\theta_{\rm{min}}$ = 41~mrad and $\theta_{\rm{max}}$ = 67~mrad, where the shower is fully contained
inside the calorimeter. 
The fiducial volume of LumiCal is thus defined to be the polar angular range 
\begin{equation}{
41 < \theta < 67 ~ \mathrm{mrad},
}\label{fiducialVlomueEQ} \end{equation}
as indicated by the dashed lines in 
Figure~\ref{fig:bhabhaXSFIG1}.
Fiducial
cuts on the minimal 
and maximal reconstructed polar angles of the particles
used for the luminosity measurement
reject
events with shower leakage through the edges of  
LumiCal. 
For
electron showers located inside the fiducial volume of LumiCal,
the energy resolution is estimated to be  
${\rm{a}}_{\rm{res}} = (0.21 \pm 0.02)~ \sqrt{\mathrm{GeV}}$. No dependence on the electron energy is found in the
energy range from 50 to 300 GeV.
In order to determine the energy of showering particles, 
the 
integrated deposited energy in the detector has to be multiplied by a calibration factor. 
The
calibration factor is found to be constant in the same energy range. 

\begin{figure}[htp]
\begin{center}
\subfigure[]{\label{fig:electronicSignalFIG1}
\includegraphics[trim=0mm 15mm 0mm 25mm,clip,width=.49\textwidth]{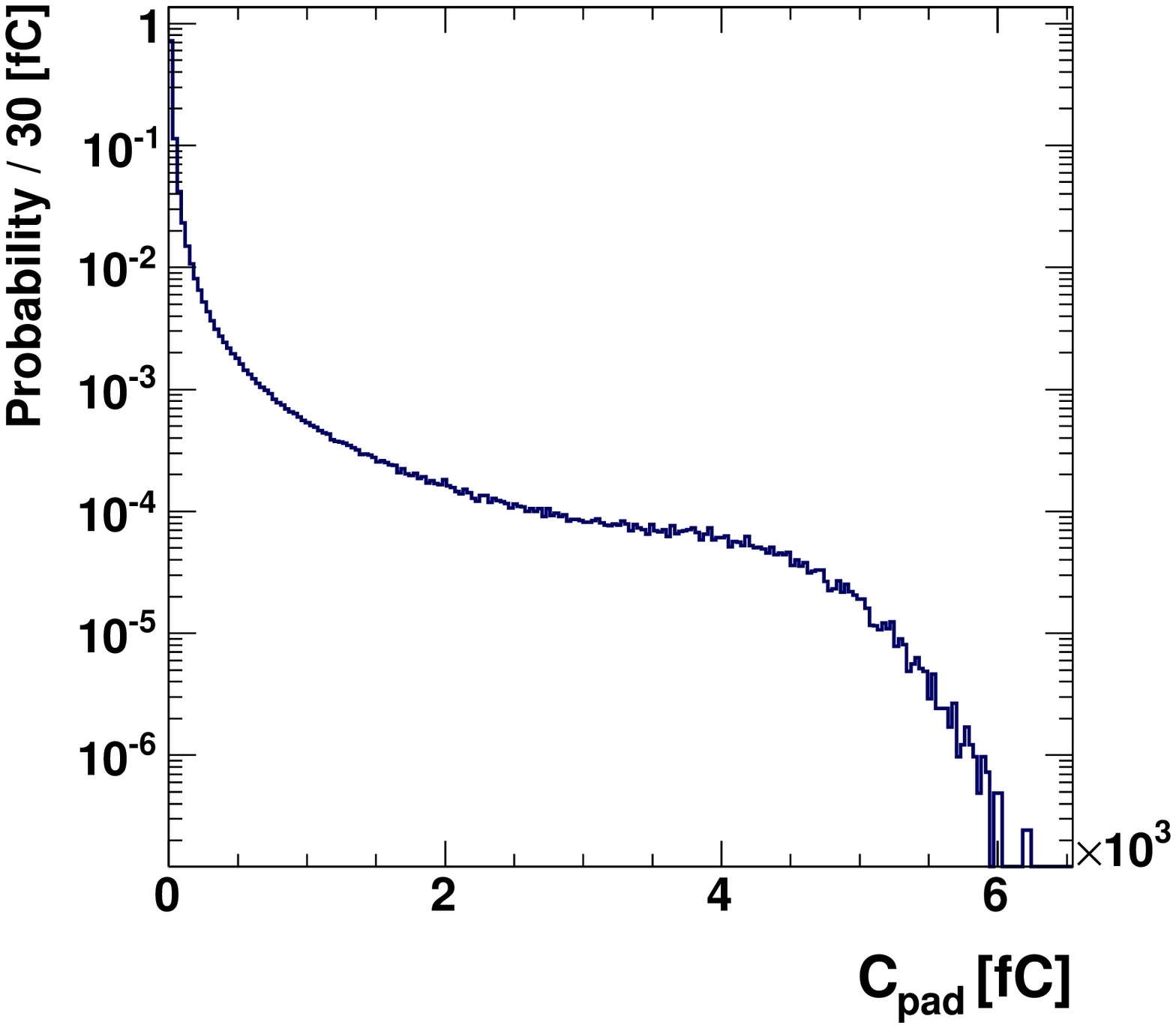}}
\subfigure[]{\label{fig:electronicSignalFIG2}
\includegraphics[trim=0mm 15mm 0mm 25mm,clip,width=.49\textwidth]{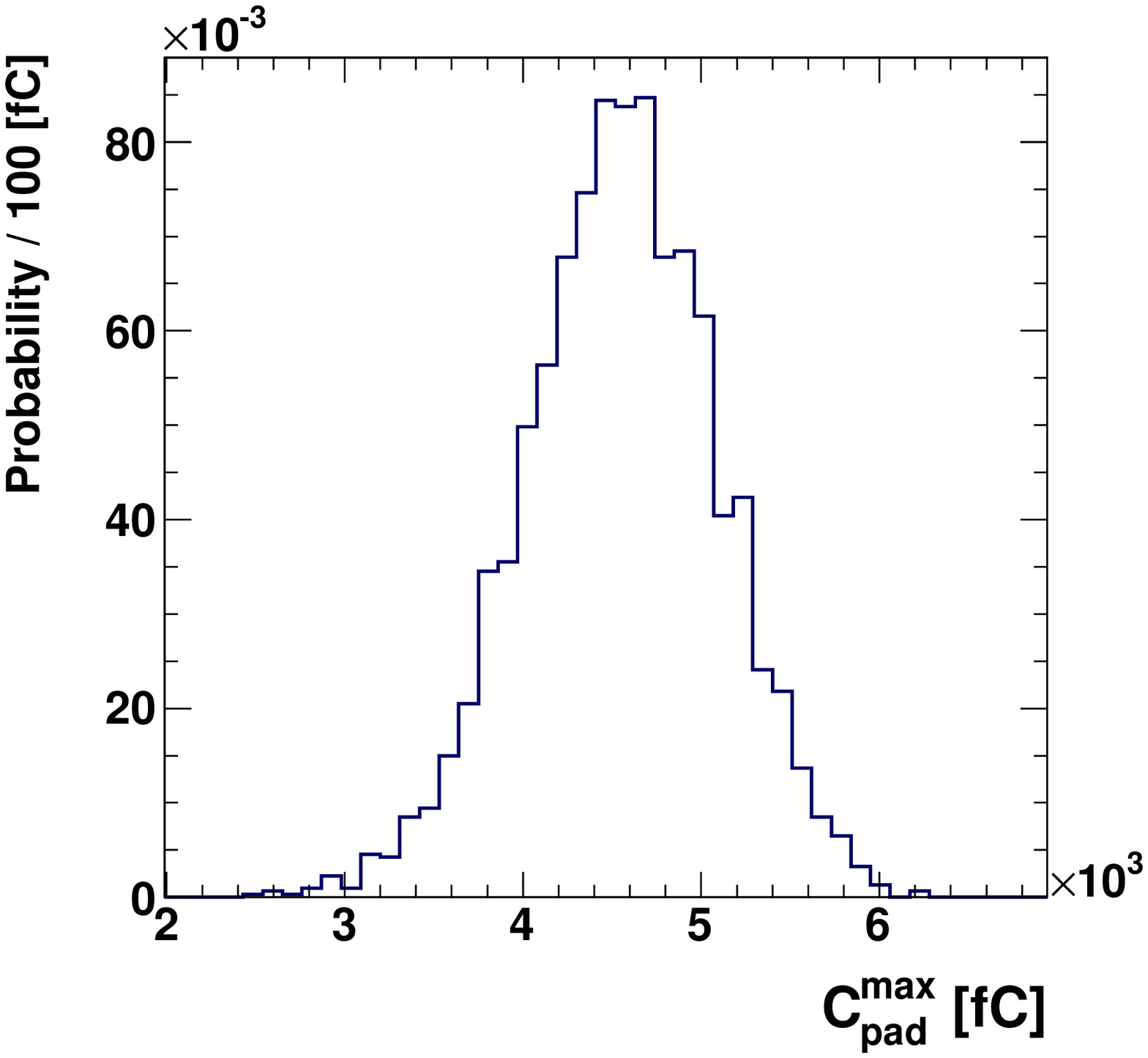}}
\caption{\Subref{fig:electronicSignalFIG1} Normalised
distribution of the charge deposited in a detector pad, ${\rm{C}}_{\rm{pad}}$, by 250~GeV electron
showers. \Subref{fig:electronicSignalFIG2} Normalised distribution of the maximal charge
collected in a single pad per shower, ${\rm{C}}_{\rm{pad}}^{\rm{max}}$, for 250~GeV electron showers.}
\end{center}
\end{figure}

The expected range of energy depositions
in the pads has been studied
for the passage of minimum ionising
particles, hereafter denoted as MIPs, and for showers of 250~GeV electrons~\cite{bib25}. 
The energy deposition in silicon is converted to released ionisation charge.
The distribution of the charge in a single pad, $\rm{C}_{\rm{pad}}$, is 
shown in Figure~\ref{fig:electronicSignalFIG1}.
It ranges between
$4 <{\rm{C}}_{\rm{pad}} < 6 \times 10^{3}$~fC.
The 
distribution of the maximal charge collected in a single pad
is shown in  Figure~\ref{fig:electronicSignalFIG2}.
About 95~\% of electron shower signals are less than $5.4\times 10^{3}$~fC.

The impact of the digitisation of the detector signal on the LumiCal performance is investigated 
in  Ref.~\cite{bib25}.
It is shown that an ADC with 8~bit resolution is sufficient to keep the energy resolution 
quoted above. No bias in the energy measurement is found. 

\subsection{BeamCal simulation studies}

BeamCal will be hit after each bunch-crossing 
by a large amount of beamstrahlung pairs. 
Their number, energy and spatial distribution 
depend
on the beam parameters and the magnetic field 
inside the detector. 
For
the nominal ILC beam-parameter set~\cite{nominal_set},
beamstrahlung pairs are generated with the GUINEA-PIG program~\cite{guinea}. 
Inside the ILC detector an
anti-DID field~\cite{anti-DID} is assumed. Beamstrahlung pairs are simulated 
in the detector, using a program based on 
GEANT4.

\begin{figure}
\subfigure[]{
\label{fig:beam_deposits}    
\includegraphics[width=0.45\columnwidth,height=0.45\columnwidth]{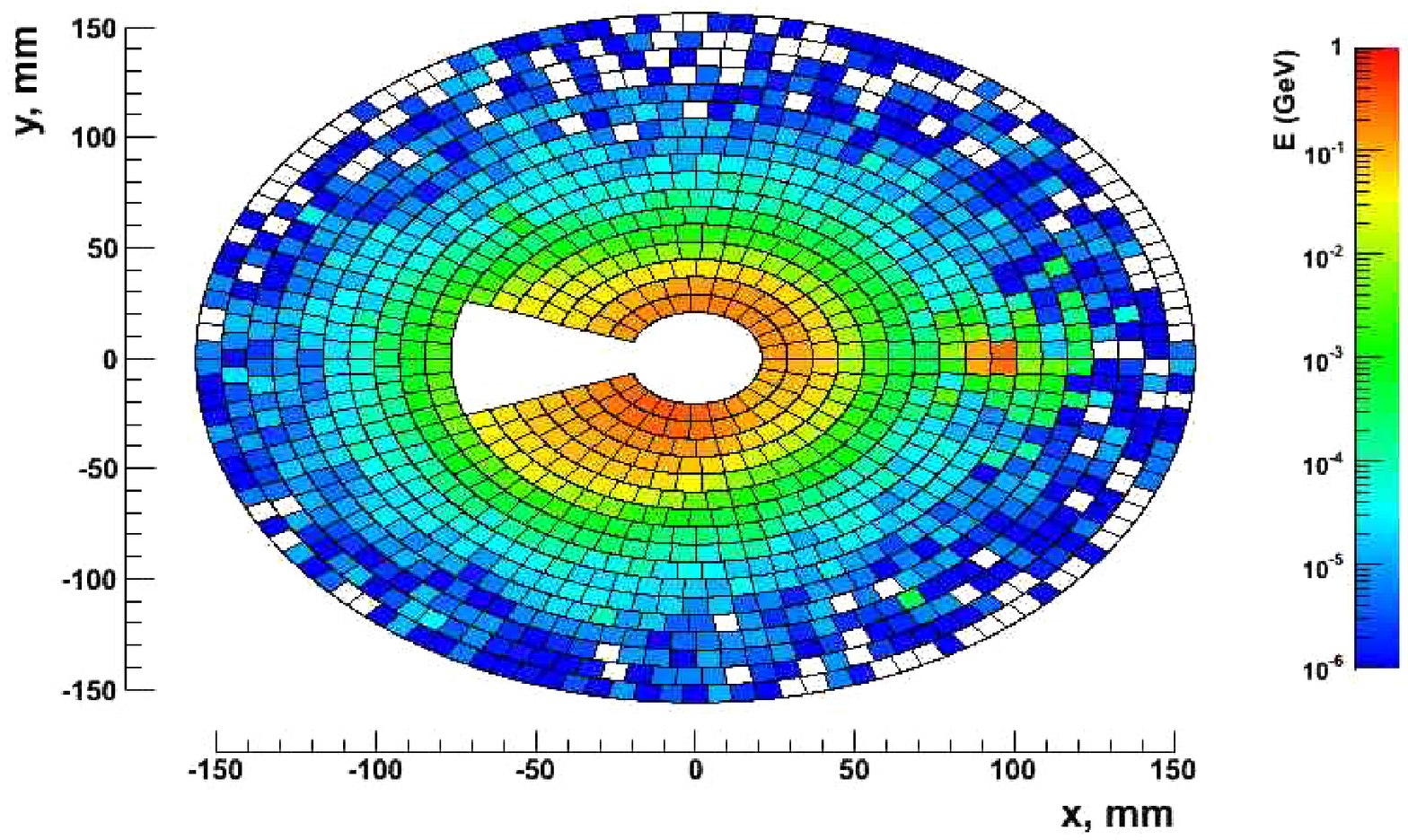}}
\subfigure[]{
\label{fig:efficiency}
\vspace{-1cm}
\includegraphics[width=0.55\columnwidth,height=0.45\columnwidth]{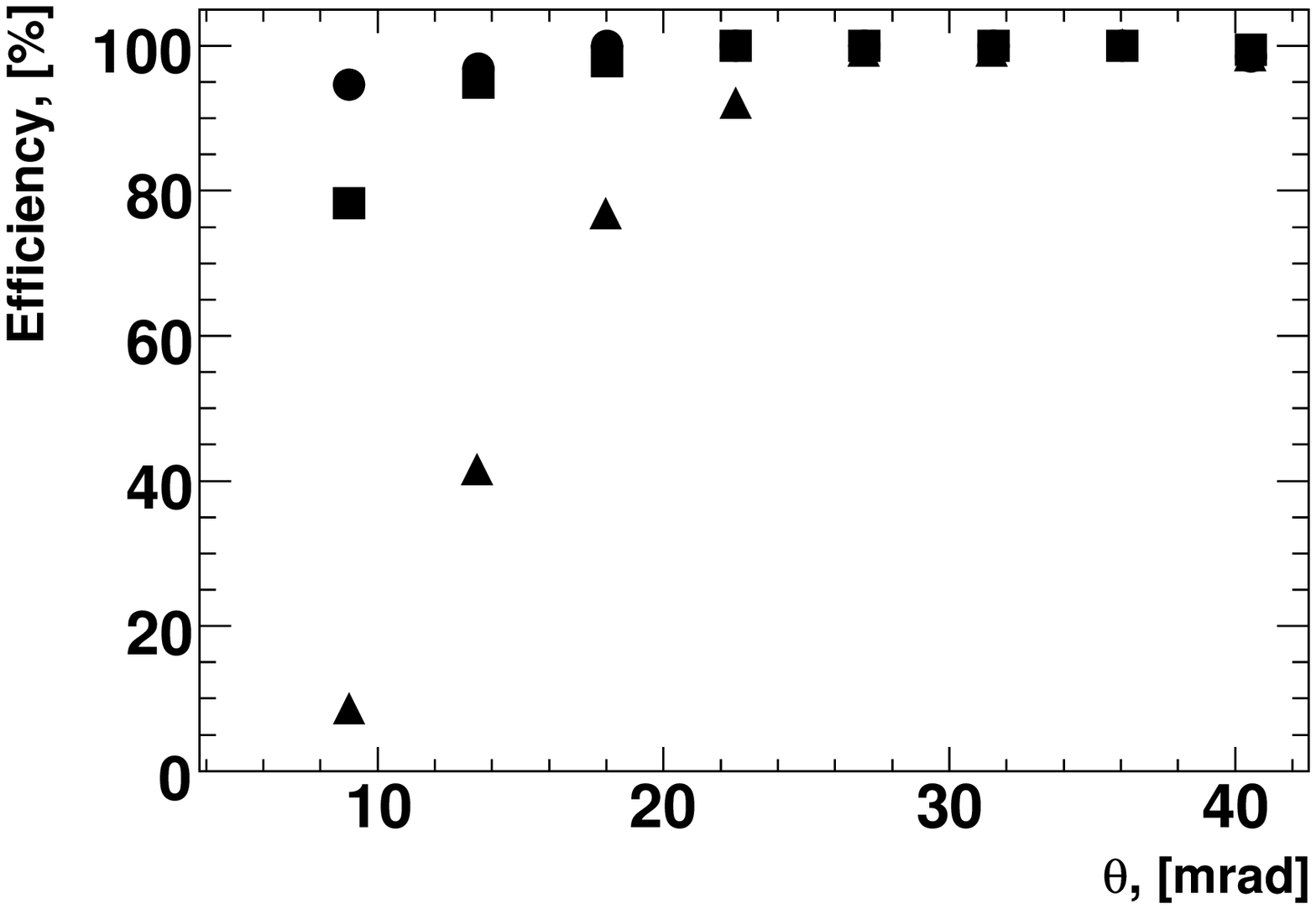}}
\caption{\Subref{fig:beam_deposits}
 The distribution of the energy deposited by beamstrahlung pairs after one bunch
crossing in the sensors of BeamCal. The depositions are integrated over pads of 7.65$\times$7.65 mm$^2$ area. 
Superimposed is the deposition of
a single high energy electron (red spot on the right side). The white area in the centre 
allows space for the beam-pipes. 
\Subref{fig:efficiency} The efficiency to detect single high energy electrons
on top of the beamstrahlung background for electron energies of 75 (triangles), 150
(squares) and 250 (circles) GeV.    
}
\end{figure}

The energy 
deposited in the sensors
of BeamCal per bunch crossing, 
about 150 GeV as shown in Figure~\ref{fig:beam_deposits},
and the shape of 
these depositions
allow a bunch-by-bunch 
luminosity estimate and the determination of beam parameters~\cite{grah1}.
From the spatial distribution of the deposited energy a set of observables, e.g. 
radial and angular moments and asymmetries, is defined. These observables are related to 
beam parameters like bunch sizes, emittances and bunch offsets by a matrix equation.
In the single parameter determination
accuracies better than 10\%~\cite{grah1} are obtained. In the multiparameter mode correlations 
appear. However, reasonable precision can still be obtained by using information from other
diagnostics devices.

For search experiments
it is important 
to detect   
single high energy 
electrons on top of the
wider spread beamstrahlung pairs.
Superimposed on the pair depositions 
in Figure~\ref{fig:beam_deposits}
is the deposition of an electron of 250 GeV, seen as 
the red spot 
on
the right side.
By performing
an appropriate subtraction of the pair deposits
and a shower-finding algorithm which
takes into account the longitudinal shower profile, 
high energy electrons 
can be detected with high efficiency, as shown in Figure~\ref{fig:efficiency}. 
This feature allows to suppress the
background from two-photon processes in a search e.g. 
for super-symmetric tau-leptons~\cite{drugakov} in a large fraction of the parameter space. 
\begin{figure}
\subfigure[]{\label{fig:largest_dose}
     \includegraphics[width=0.45\columnwidth,height=6.5cm]{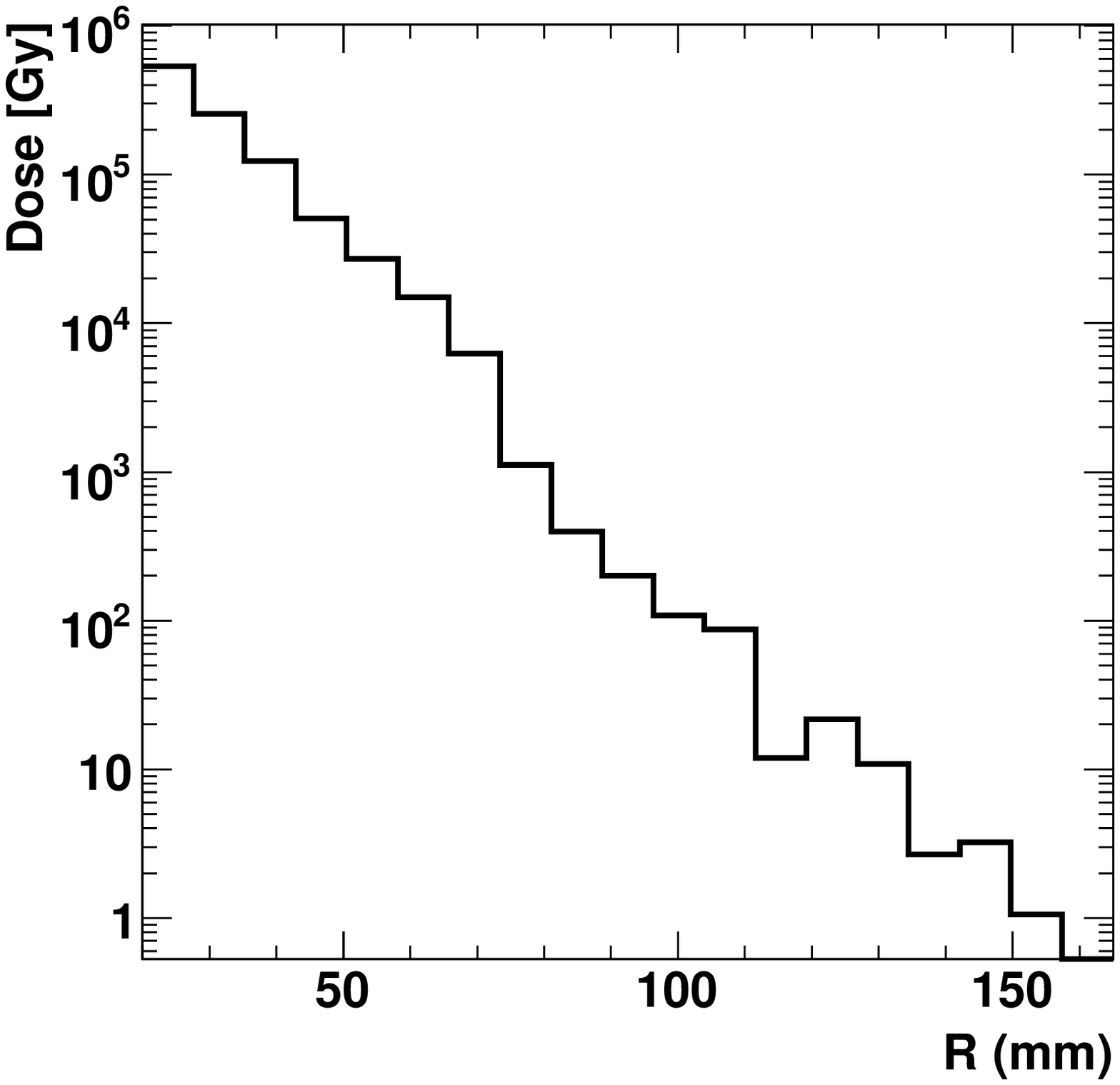}
}
\subfigure[]{
\label{fig:neutron_fluxb}    
\includegraphics[width=0.45\columnwidth,height=6.5cm]{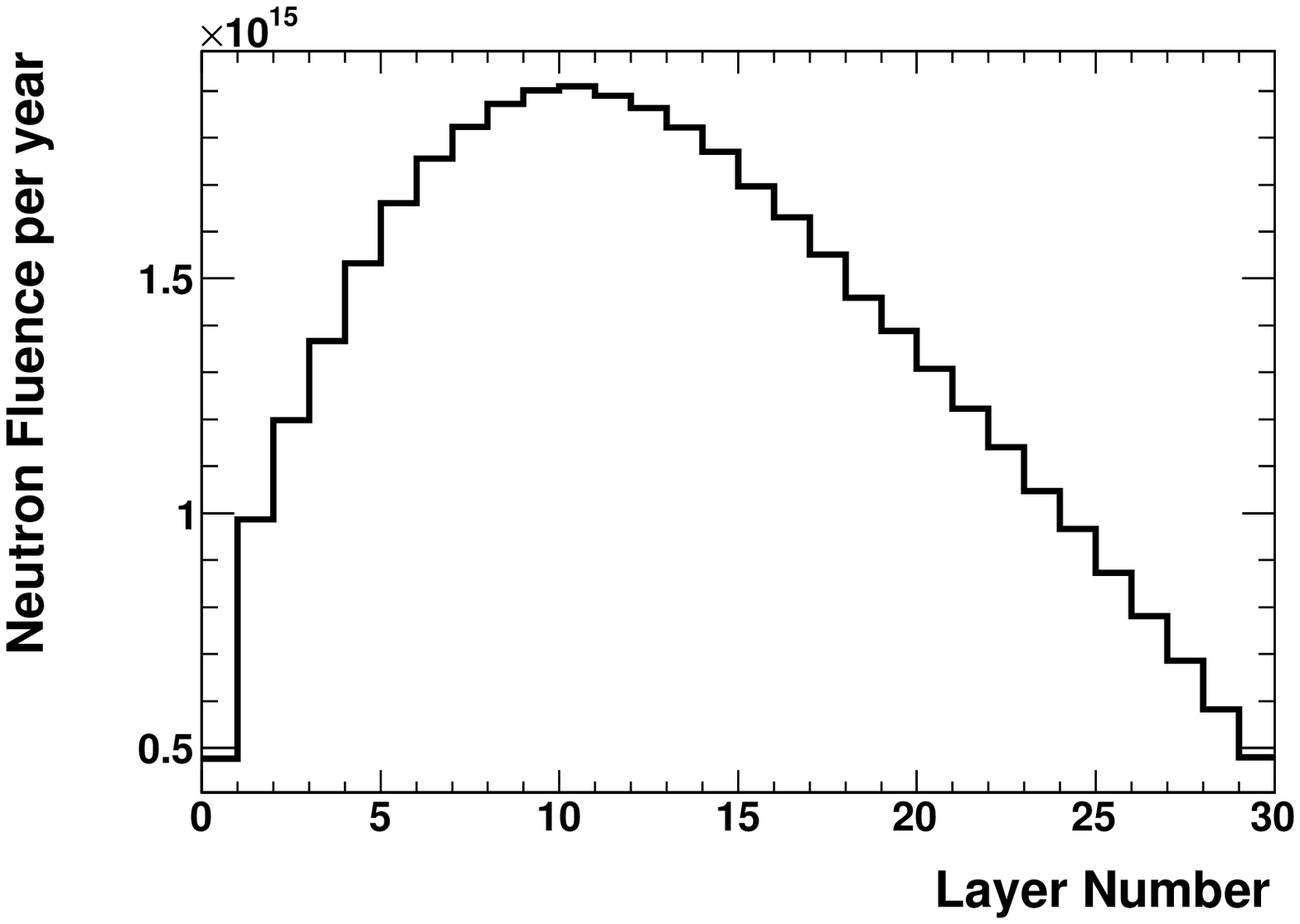}}
\caption{\Subref{fig:largest_dose}
The dose in BeamCal 
sensors per year as a function of the radial distance from the beam.
\Subref{fig:neutron_fluxb} The fluence of neutrons per year inside the sensors of BeamCal
as a function of the sensor layer number using the cascade model of Bertini. An integrated luminosity of
500 fb$^{-1}$} is assumed.
\end{figure}

The range of signals expected on the pads
was estimated. Including the depositions from beamstrahlung signals up to 40 pC
are expected. Digitising the signals with an ADC with 10 bit resolution
has no impact on the performance of the calorimeter.

GEANT4 simulations are also used to determine the expected dose
and the neutron fluence in the sensors after one year of operation with nominal
beam parameters.
The dose in a sensor layer at the depths of the shower maximum as a function of 
the radius 
is shown in  Figure~\ref{fig:largest_dose}.
In the innermost ring of the calorimeter a dose of about 0.5 MGy is expected. Since the dose is non-uniformly
distributed as a function of the azimuthal
angle, it approaches 1 MGy per year  in some sensor areas of the inner rings.

\begin{figure}
\begin{center}
\includegraphics[width=0.4\columnwidth]{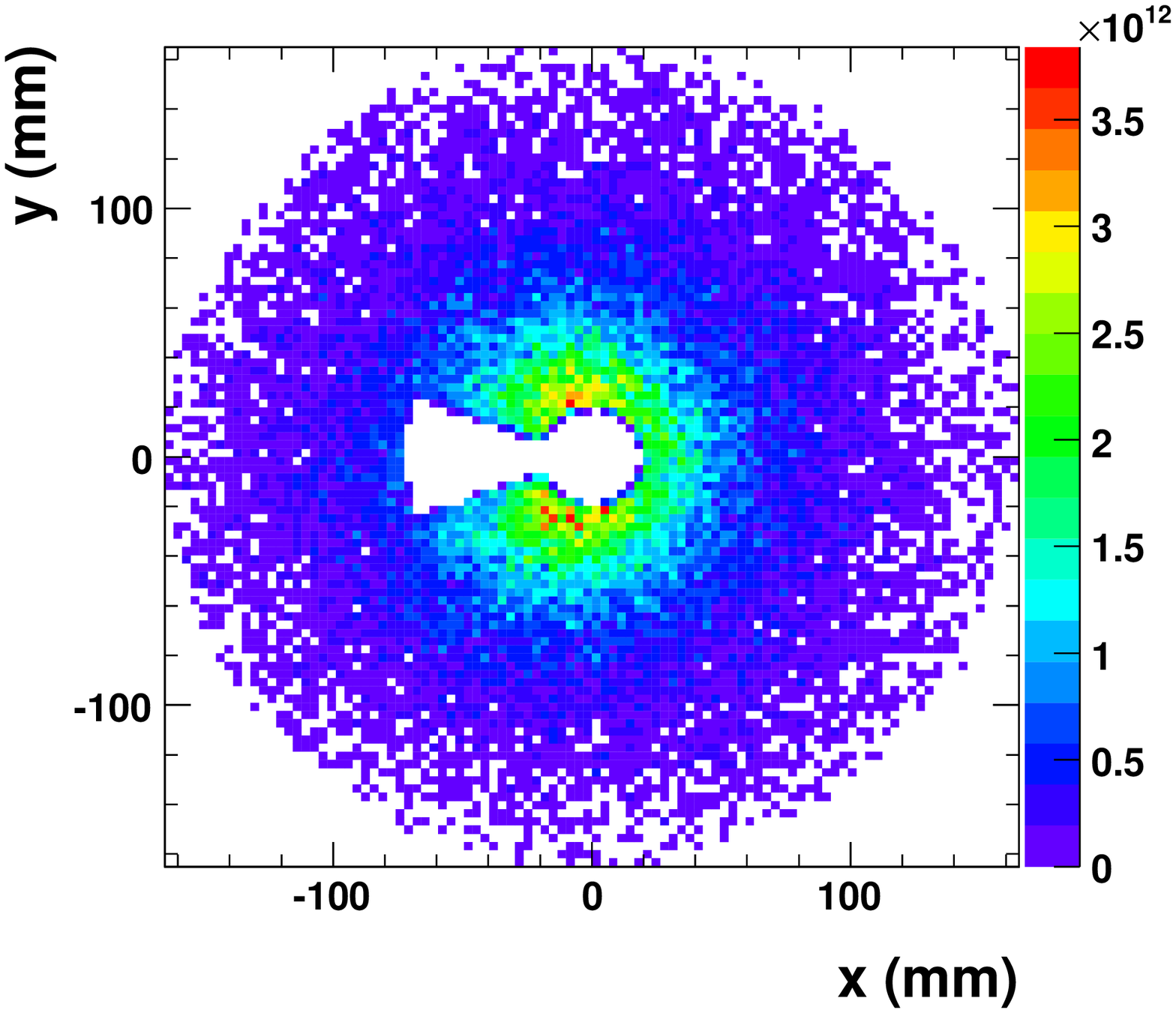}
\caption[]{\label{fig:neutron_flux_rad} 
The fluence of neutrons per $\rm{mm^2}$ and year crossing a sensor of BeamCal
near the shower maximum
using the cascade model of Bertini.
}
\end{center}
\end{figure}

The neutron fluence is estimated using in GEANT4
the cascade 
model of Bertini~\cite{bertini}.
The fluence per year of running at 
nominal beam parameters is shown in Figure~\ref{fig:neutron_fluxb}
as a function of the sensor layer number. Fluences up to 2$\times$10$^{15}$ per layer are expected
near the shower maximum. 
Other GEANT4 models predict lower neutron fluences, particularly
at low neutron energies~\cite{RomJourPhys}.
The distribution of the fluence of neutrons  
in the sensor 
layer with the maximum fluence is 
shown in Figure~\ref{fig:neutron_flux_rad}.  
With
the cascade model of Bertini,
a neutron fluence of 0.4$\times$ 10$^{12}$ neutrons per mm$^2$ and year  
is expected near the beam-pipe. Albeit this is still an order of magnitude
less than predicted for LHC detectors 
near the beam pipe dedicated tests of sensors are planned.

\subsection{Pair monitor simulations} 

Additional and independent information on beam parameters will 
be obtained from the pair monitor~\cite{tauchi1, tauchi2}.
The device will consist of one layer of 
silicon pixel sensors, with pixel size of 400$\times$400 $\mu$m$^2$, 
just in front of BeamCal
to
measure the 
number density
distribution of beamstrahlung pairs.
Here
we investigated the sensitivity to the horizontal and vertical 
bunch sizes, $\sigma_{\mathrm{x}}$ and $\sigma_{\mathrm{y}}$, and the ratio 
of the vertical displacement between bunches crossing 
to their vertical size, $\Delta_{\mathrm{y}}$.

\begin{figure}[hbt]
\begin{center}
\includegraphics*[width=15cm]{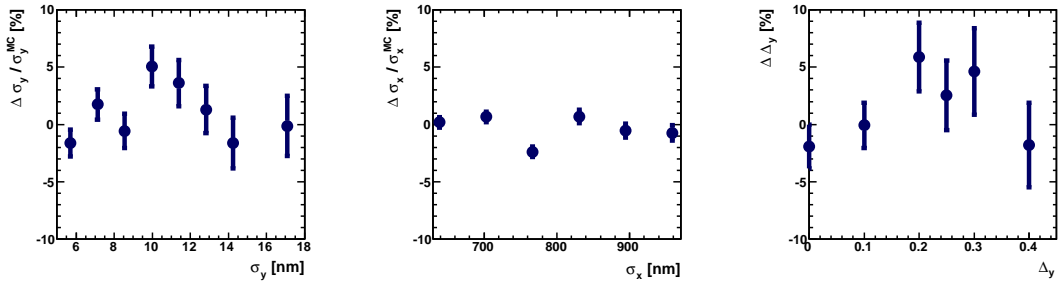}
\end{center}
\caption{\label{fig:reso}
The relative deviations of the vertical, $\sigma_y$, and horizontal, $\sigma_x$, beam sizes, and
the ratio of vertical displacement to the vertical beam size, $\Delta_y$, averaged over 50 bunch crossings 
measured by the pair monitor.}
\end{figure}

To reconstruct the beam profile several observables characterising the
number density of pairs at the front 
face of BeamCal are used~\cite{ito}.
Bunch crossings are simulated for certain ranges of  
$\sigma_{\mathrm{x}}$, $\sigma_{\mathrm{y}}$ and $\Delta_{\mathrm{y}}$, and 
each of these observables is fitted with a second order polynomial.
Then, several thousand bunch crossings are generated using different sets of beam parameters
and $\sigma_{\mathrm{x}}$, $\sigma_{\mathrm{y}}$, and $\Delta_{\mathrm{y}}$ are 
reconstructed with the inverse matrix method. 
Figure \ref{fig:reso} shows a few examples of the results displayed as   
the difference between the beam parameters reconstructed and set 
in the simulation divided by the latter, averaged over 50 bunch crossings.     
These quantities are compatible with zero.  The relative uncertainties, averaged
over about 100 such reconstructions of 
vertical and horizontal beam 
sizes and the relative vertical displacement are 10.1\%, 3.2\% and  8.0\%, respectively.

\section{Mechanical concepts}

\begin{figure}[htpb]
\centering
\subfigure[]{\label{fig:beamcal123}
\includegraphics[width=0.4\columnwidth]{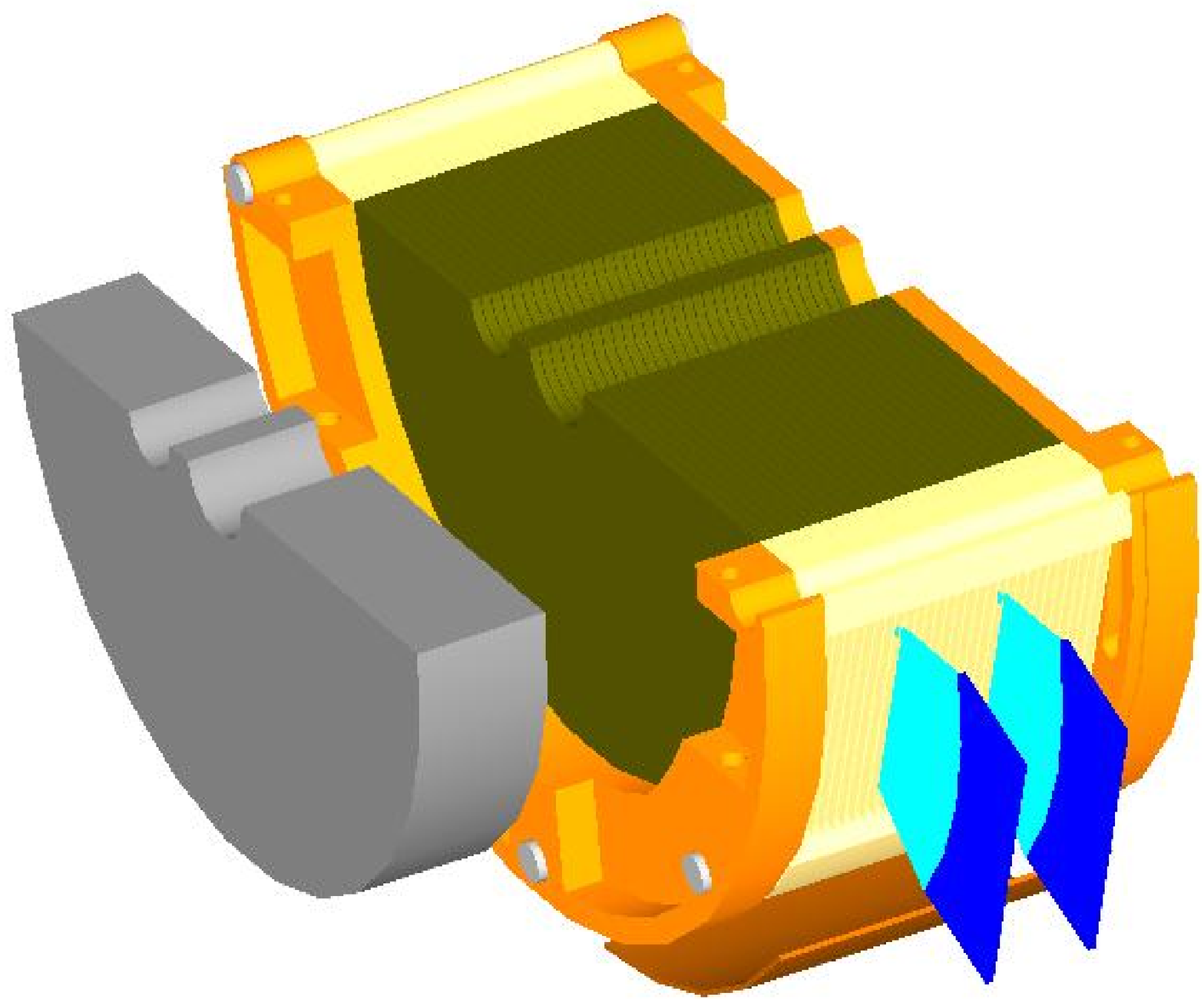}
}
\subfigure[]{\label{fig:beamcal124}
\includegraphics[width=0.4\columnwidth]{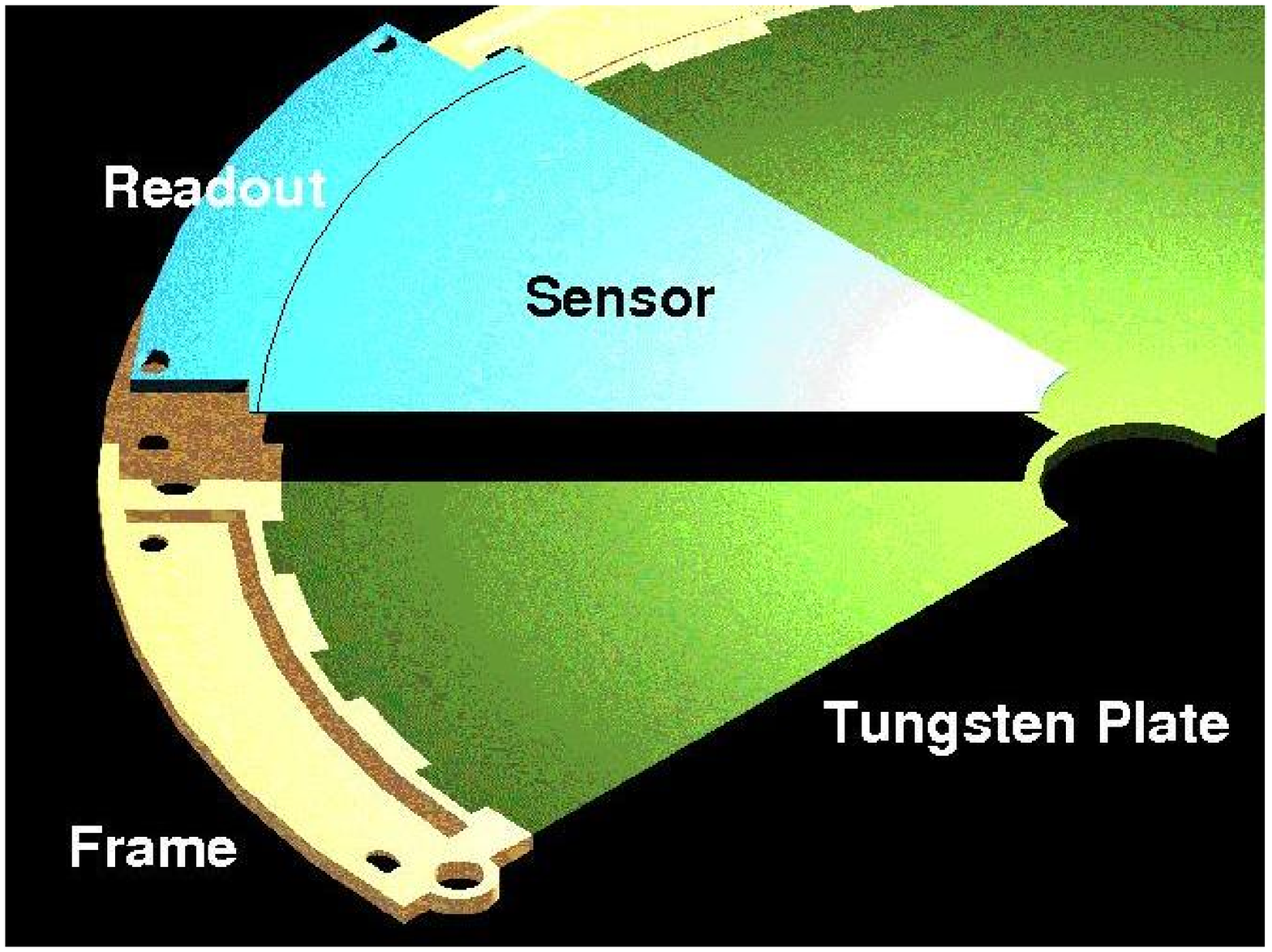}
}
\caption{\Subref{fig:beamcal123} A half-cylinder of BeamCal. The brown block is the 
tungsten absorber structure interspersed with sensor layers. 
The orange structure represents the mechanical frame. The blue segments
at the outer radius indicate the front-end electronics.
In front of the calorimeter a graphite shield, shown in grey, reduces the amount of low energy particles 
back-scattered into the tracking detectors.
\Subref{fig:beamcal124} 
A half-layer of an absorber disk assembled with a sensor sector and the front-end
readout.}
\end{figure}

On the basis of the simulation results mechanical designs of both calorimeters 
are developed.
To allow their installation after the beam-pipe is in place, both calorimeters
consist of two half-cylinders.
A schematic of a half cylinder of BeamCal is shown in Figure~\ref{fig:beamcal123}. 
The tungsten absorber disks are embedded in a mechanical frame stabilised by 
steel rods. Each layer is composed of a tungsten half-disc surrounded by a brass 
half-ring as shown in Figure~\ref{fig:beamcal124}. Precise
holes in the brass ring will ensure a position accuracy of better than 100$\mu$m.
The sensors are fixed on the tungsten
and connected via a flexible PCB to the front-end readout.
The distance between two adjacent tungsten plates is kept 
to
1~mm
to approach the smallest possible
Moli\`{e}re radius.  
The sensors of BeamCal are structured into pads of about 8$\times$8 mm$^2$ size
allowing the maximum electron detection efficiency~\cite{elagin_snowmass}. 
Due to the required high radiation tolerance,
GaAs sensors are foreseen. For the innermost part of BeamCal, adjacent to the beam-pipes, also
CVD\footnote{Chemical Vapour Deposition} diamond is considered.

\begin{figure}
\centerline{\includegraphics[width=0.4\columnwidth]{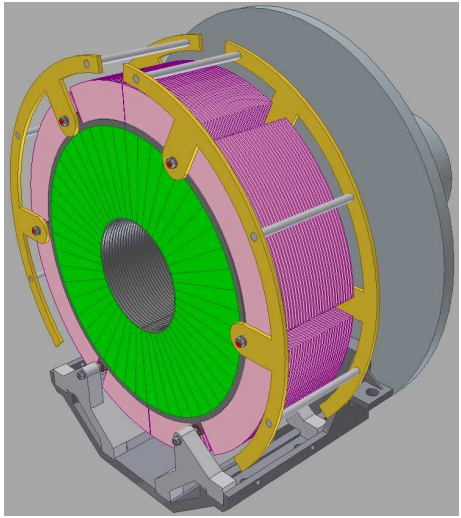}}
\caption{The mechanical structure of LumiCal. Tungsten disks are
precisely positioned using 4 bolts which are stabilised by additional steel 
rings on both sides of the cylinder. 
}\label{fig:lumical}
\end{figure}

The design of LumiCal is similar~\cite{LumiCal_mechanics}. Since it is a precision device,
special care is devoted to the mechanical stability and position control. 
The tungsten half-discs are held by special bolts. 
For a half barrel structure as shown in Figure~\ref{fig:lumical} a finite element 
simulation is performed. The calorimeter weight leads to a maximal vertical displacement
of 20 $\mu$m. For a temperature difference of 
1~K
over a disk, the deformation of the shape of the tungsten plate is estimated 
to be 25 $\mu$m. 
To match the 
requirements on the precision of the lower polar angle measurement, the 
sensor positions at the inner acceptance radius must be controlled to better than 40 $\mu$m. 
Other critical quantities are 
the distance between the two calorimeters
and the position of the beam with respect to the
calorimeter axis. The former must be known to about 1 mm and the latter to 500 $\mu$m.
A laser based position monitoring system has been developed~\cite{laser_alignment} to control the position
of LumiCal 
over short distances with
$\mu$m precision.

For LumiCal, sensors made of high-ohmic n-type silicon are foreseen. The thickness of the sensors is 
about 300 $\mu$m. The p$^+$ side is segmented in polar and azimuthal pads and the backside is fully 
metallised. To keep the Moli\`{e}re radius small the gap for the sensors is 1~mm. 
The signals on the pads of both calorimeters are led by thin copper strips on a 
Kapton foil to 
the front-end electronics positioned at the outer radius of the calorimeter.

%% file: sys_eff.tex
\section{Systematic effects in the luminosity measurement}

Several phenomena which may have an impact on the 
luminosity measurement are considered. 
These are: pinch effect and beamstrahlung, 
background from two-photon processes, the 
resolution and scale of the electron energy measurement and the beam polarisation.

\subsection{Pinch effect and beamstrahlung}

Due to the pinch effect the luminosity for given bunch charges and sizes
will be enhanced. However, 
electrons and positrons
may radiate photons prior to Bhabha scattering.
In addition, final state particles are 
deflected inside the bunch.
The result is
a reduction of the Bhabha event counting rate in a given range of low polar angles. 
The reduction is found to depend on the selection criteria for Bhabha events. For a selection optimised for
nominal ILC beam parameters at 500 GeV centre-of-mass energy, it amounts to 
1.51$\pm$0.05\%~\cite{CECILE},
where the quoted
uncertainty stems from the statistics in the simulation.
The dominant contribution to the loss is due 
to the reduction in the centre-of-mass energy caused by beamstrahlung. The latter leads to
an effective centre-of-mass energy distribution called luminosity spectrum.

In the measurement of the luminosity, the loss of Bhabha events has to be corrected for.
The impact of beamstrahlung can be estimated from the measured 
luminosity spectrum with a relative uncertainty 
of about 10$^{-3}$~\cite{CECILE}. The impact of the deflection 
inside the bunch depends mainly on the horizontal bunch-size, $\sigma_{\rm{x}}$, and the
bunch length, $\sigma_{\rm{z}}$. Assuming that one can control these two quantities with a 
relative uncertainty of 5\%\footnote{In Ref.~\cite{grah1} the estimated uncertainty of e.g. $\sigma_{\rm{x}}$
varies between 0.5\% and 6.5\%, depending on the number of free 
beam parameters in the analysis. A similar range of precision is obtained for $\sigma_{\rm{z}}$.},
the uncertainty of a correction 
to the luminosity is
about 1.5$\times$ 10$^{-3}$~\cite{CECILE}.   
\begin{figure}[htb!]
\centerline{\includegraphics[width=0.25\columnwidth,height=4.cm]{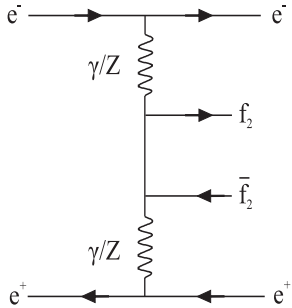}}
\caption{The Feynman graph for the dominant process in four-fermion production.}\label{fig:Multiperipheral}
\end{figure}

\subsection {Background from four-fermion production}

Four-fermion production is known to have a large cross section 
with maxima at low polar angles. 

It is dominated by the diagram shown in Figure~\ref{fig:Multiperipheral},
where two virtual photons are exchanged between electron spectators. 
We used the WHIZARD~\cite{WHIZARD} event generator to obtain 
samples of 
events for final states with leptons in the inner legs. The generator was tuned to experimental data
of the process e$^+$e$^- \rightarrow$ e$^+$e$^-$ c$\bar{\rm{c}}$ using data from LEP and other 
accelerators~\cite{whizard_tune}. 
The cross-section of four--lepton production amounts to 12.0$\pm$0.5 nb at 500 GeV
when the momenta of the exchanged photons are required to be larger than 0.1 GeV/c.
The spectators 
remain at high energy. Less than 1\%  
of them hit the 
luminosity calorimeter and become a background for Bhabha events.
A Bhabha event sample has been generated with a cross-section of 4.70$\pm$0.03 nb 
at 500 GeV centre-of-mass energy, 
using the BHLUMI \cite{BHLUMI} 
event generator. The LumiCal response is simulated using BARBIE V4.3 \cite{BARBIE}, a GEANT3 based 
simulation program.
The following event selection criteria are
applied: the polar angle of the reconstructed shower must be within
the LumiCal fiducial volume at one side and
within
 $\theta_{\rm{min}}+4$ mrad and 
$\theta_{\rm{max}}$ -- 7 mrad on the other. In addition, the total energy deposited in both calorimeters
must be more than 80$\%$ of the center-of-mass 
energy. 
These criteria are optimised to reduce the impact of beamstrahlung and deflection
on the Bhabha event counting to the amount given 
in the previous section~\cite{CECILE}.
The selection efficiency of Bhabha scattering events is about 68$\%$.

\begin{figure}
\centerline{    
\includegraphics[width=\columnwidth]{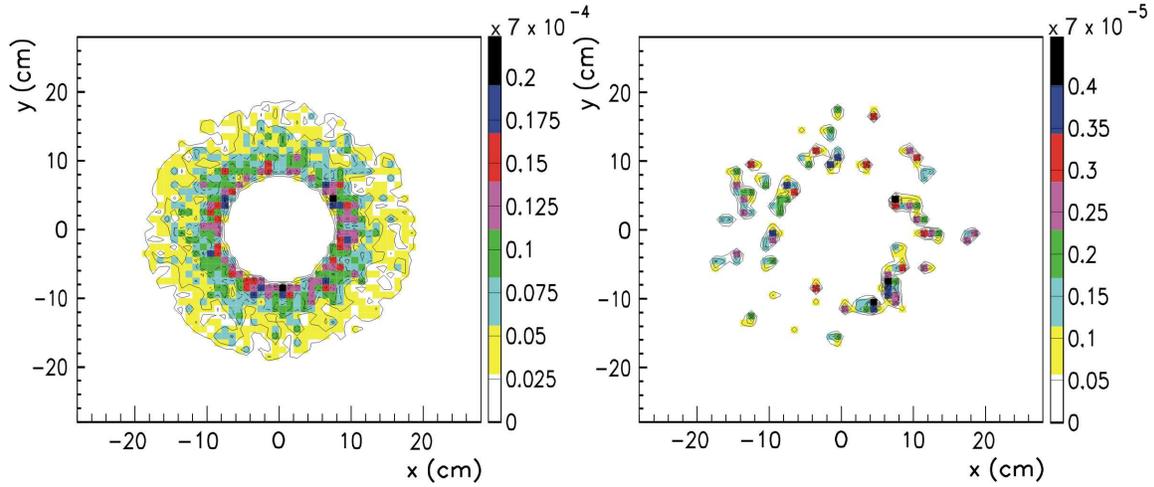}}
\caption []{\label{fig:HITS}
Average number of hits originating from four-fermion 
interactions per bunch crossing 
on the first plane of LumiCal 
at 500 GeV, before (left) and after (right) application of Bhabha event selection criteria.   }
\end{figure}

Four-fermion events in the LumiCal are to a large fraction rejected by the Bhabha selection
criteria. This is illustrated in Figure~\ref{fig:HITS} where the hits of 
particles from the 
four-fermion final states in 
the front plane of LumiCal per bunch crossing are shown before and after applying the 
Bhabha event selection. 
The fraction of four-fermion final states in the selected Bhabha event 
sample is 2.3$\times$10$^{-3}$.

At LEP energies agreement between measurements and modelling of four-fermion processes was
obtained within 20\%~\cite{whizard_tune}. 
Assuming that at 500 GeV it will be possible to model these processes with a precision of 
40\%, correcting the luminosity measurement
correspondingly will lead to an uncertainty of 0.9$\times$10$^{-3}$.

\subsection{Effects of a bias in the energy resolution and the energy scale}

One of the criteria to select Bhabha events is the total energy measured in the calorimeters,
required
to be larger than 80\% of the centre-of-mass energy. A possible bias in the energy resolution
or the energy calibration will result in a change of the number of selected Bhabha events and 
hence in the measured luminosity.

\begin{figure}[htp]
\begin{center}
\subfigure[]{\label{fig:scale}
\includegraphics[width=0.44\columnwidth,height=4.5cm]{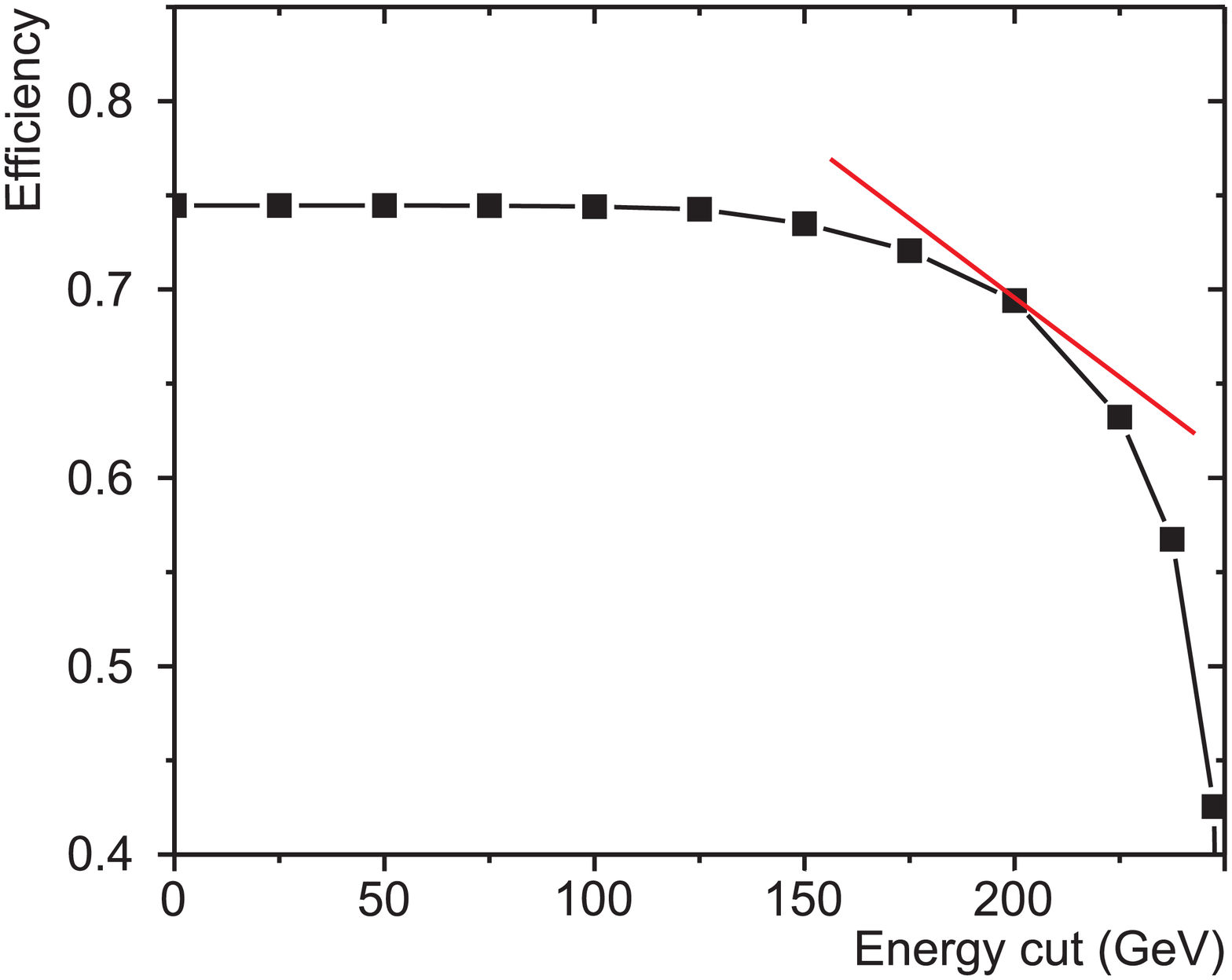}} 
\subfigure[]{\label{fig:resolution}
\includegraphics[width=0.52\columnwidth]{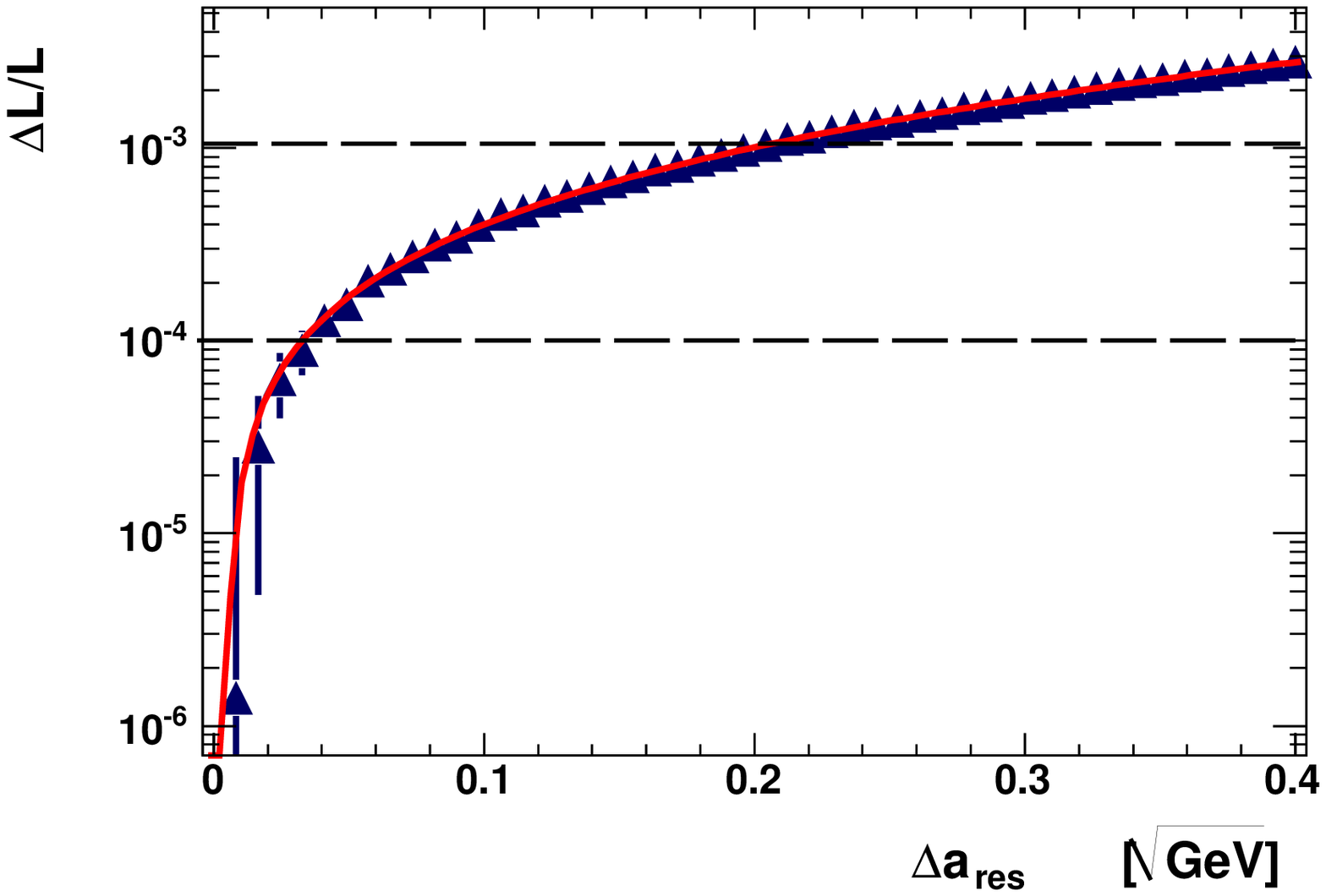}} 
\end{center}
\caption{\Subref{fig:scale} The selection efficiency for Bhabha events as a function of the measured shower energy,
\Subref{fig:resolution} the shift of the measured luminosity as a 
function of the bias in the energy resolution parameter ${\rm{a}}_{\rm{res}}$.
}
\end{figure}

The selection efficiency for Bhabha events as a function of the required energy in the calorimeters
is shown in Figure~\ref{fig:scale}.
At the position of the cut in the measured calorimeter energy the slope of the tangent to the function is
about --1.8$\times10^{-3}$.
To keep the shift of the luminosity below $10^{-3}$, the cut in the measured calorimeter 
energy must be controlled with a precision of about 400 MeV.
A study done allowing a constant offset in the measured energy leads to a 
similar requirement~\cite{e_offset}. 

The effect of a bias in the energy resolution, ${\rm{a}}_{\rm{res}}$ in eqn.~\ref{engyResEQ}, 
is illustrated in 
 Figure~\ref{fig:resolution}. 
We estimate that if ${\rm{a}}_{\rm{res}}$ can be controlled within
20\%, it will contribute to the luminosity uncertainty by about $10^{-4}$ .

\subsection{Impact of electron and positron polarisation}
To exploit the full physics potential of a linear collider, 
electron and positron beams will be polarised. Polarisation will also change the Bhabha
cross section in the acceptance range of LumiCal up to a few per cent~\cite{report_2007}.
In the current design the maximum values for electron and positron polarisation
are 0.8 and 0.6, respectively, with an uncertainty of 0.0025~\cite{polpaper}.
Using these values the shift in the Bhabha cross section is 2.3$\times$10$^{-2}$ with an uncertainty of
1.9$\times$10$^{-4}$.

\subsection{Summary of systematic uncertainties}

In addition to effects studied in this chapter also the  
impact of the polar angle resolution and polar angle bias as estimated in 
section 2.1 are included. All uncertainties based on the current level of understanding
are summarised in Table~\ref{tab1}. They
are considered as being uncorrelated, 
leading currently to a total uncertainty of 2.3 $\times$ 10$^{-3}$.
\begin{table}[hbt]
\begin{center}
\caption[]{\label{tab1}}{The estimated systematic uncertainties on the luminosity measurement from all sources considered above 
at a centre-of-mass energy of 500 GeV.}
\begin{tabular}{|p{3.0cm}|p{3.0cm}|p{3.0cm}|p{3.0cm}|}
\hline
Source              & Value          &  Uncertainty         &  Luminosity Uncertainty    \\       
\hline   
$\sigma_{\theta}$    & 2.2$\times$10$^{-2}$ & 100\%          &  1.6$\times$10$^{-4}$        \\
$\Delta_{\theta}$    & 3.2$\times$10$^{-3}$ & 100\%          &  1.6$\times$10$^{-4}$        \\
 $a_{\rm{res}}$           & 0.21           & 15\%           &   10$^{-4}$                 \\
luminosity spectrum  &                    &                &        10$^{-3}$      \\
bunch sizes $\sigma_{\rm{x}}$, $\sigma_{\rm{z}}$, & 655 nm, 300 $\mu$m & 5\% &  1.5$\times$10$^{-3}$  \\
two photon events   &  2.3$\times$10$^{-3}$ & 40\%         &    0.9$\times$10$^{-3}$        \\
energy scale        &  400 MeV             & 100\%       & 10$^{-3}$                 \\
polarisation, e$^-$, e$^+$ &  0.8, 0.6            & 0.0025        & 1.9$\times$10$^{-4}$  \\
\hline
total uncertainty   &                      &               &  2.3 $\times$ 10$^{-3}$  \\
\hline
\end{tabular}

\end{center}
\end{table}
The reduction of the largest uncertainty, due to the deflections of final state electrons or positrons
inside the bunch, 
needs further
investigation.
Also the energy scale uncertainty
may be reduced
by a proper calibration.

%% file: sensors_jinst.tex
\section{Sensor development}

\subsection{Sensors for BeamCal}
\begin{figure}
\centerline{    
\includegraphics[width=0.35\columnwidth,height=8cm,angle=90]{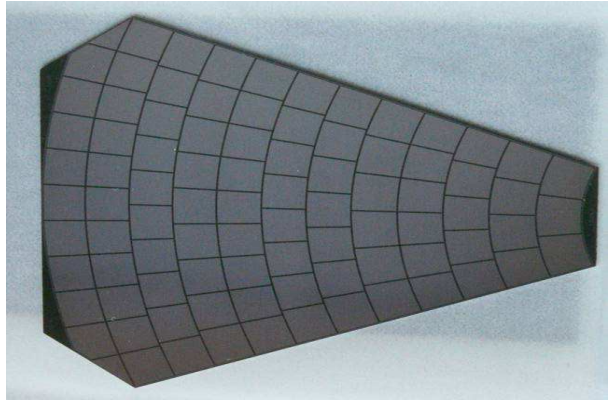}}
\caption []{\label{fig:GaAs}
A prototype of a GaAs sensor sector for BeamCal with pads of about 30 mm$^2$ area.   }
\end{figure} 
The challenge of BeamCal is to find sensors 
tolerating about one MGy of dose per year.
So far polycrystalline CVD diamond sensors of 
1 cm$^2$ size and larger sectors of GaAs pad sensors,
as shown in Figure~\ref{fig:GaAs}, have been studied.
Irradiation is done using a 10 MeV electron beam at the S-DALINAC
accelerator~\cite{sdalinac}.
The intensity is varied between 10 and 100 nA corresponding to dose rates
between 10 and 200 kGy/h.
Since
large area CVD diamond sensors are extremely expensive, 
they may be used only at the innermost part of
BeamCal. At larger radii GaAs sensors appear
to be a promising option.

\subsubsection {GaAs sensors}

Large area GaAs sensors are obtained from the Tomsk State University.
They are produced using the liquid encapsulated Czochralski method and are doped with tin and
tellur as shallow donors and chromium as a deep acceptor.

\begin{figure}[tb]
  \centering
  \includegraphics[width=0.8\textwidth]{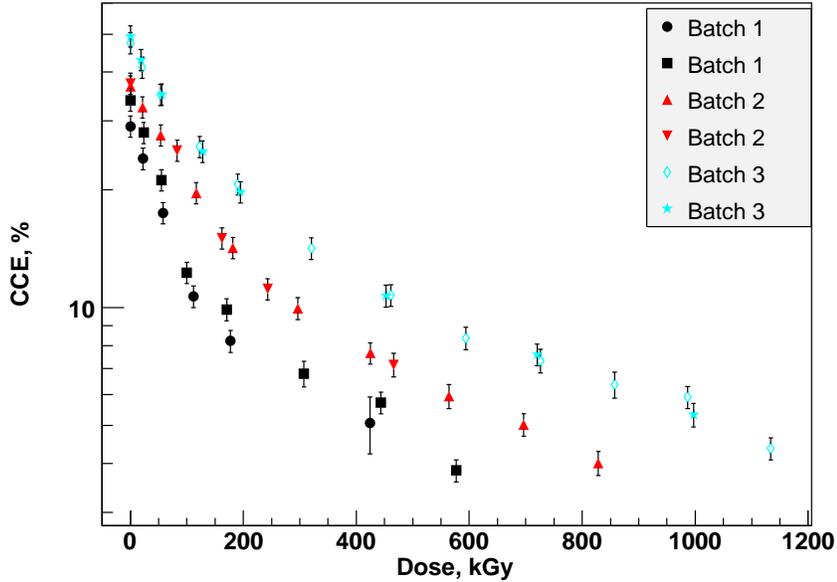}
  \caption{The CCE as a function of the absorbed dose for the GaAs sensors with
    different donor concentrations. The donor is tellur for batches 1
    and 2 and tin for batch 3.} \label{fig:gaascce}
\end{figure}

Three batches with different concentrations of dopants are irradiated up to 1.2 MGy
and the charge collection efficiency, CCE, is measured as a function of the absorbed dose. 
The results are shown in Figure~\ref{fig:gaascce}.
The charge  collection efficiency depends slightly on the dopant concentration.
The sensors with a lower donor concentration show a larger initial 
charge collection efficiency and the 
decrease of the charge collection efficiency as a function of the 
absorbed dose is less steep. 
The smallest decrease of the CCE as a function of the dose is observed for tin donor.
A MIP signal is separated from the pedestal up to a dose of 600 kGy for the sensors 
with lower donor concentration.
The leakage current of a pad at room temperature before irradiation 
is about 200 nA at an applied voltage of 50 V. After exposure of a dose of 1.2 MGy 
leakage currents of up to a factor 2 larger were found.
The pad capacitance is measured to 12 pF. 
The results are consistent with previous measurements~\cite{ieee3}.

\subsubsection{CVD diamond sensors}

For polycrystalline diamond sensor samples of 1 cm$^2$ area and 500 $\mu$m
thickness the linearity of the response and the
leakage current and the signal collection efficiency have been investigated
as a function of the absorbed dose
~\cite{ieee2}. The signal size depends linearly on the number of charged particles crossing the sensors
for up to 5$\times$10$^6$ particles in 10 ns.
The leakage current, less than 1 pA at room temperature,
depends only slightly on the absorbed dose up to 7 MGy.
The charge collection efficiency rises by a factor of two for doses between 0.5 to 1 MGy, then 
drops  
smoothly approaching the charge collection efficiency of a non-irradiated sensor. 
Provided the sensor is continuously irradiated, this efficiency is reached at about 7 MGy.

\subsection{Sensors for LumiCal}

\begin{figure}
  \centering
  \includegraphics[width=0.45\textwidth]{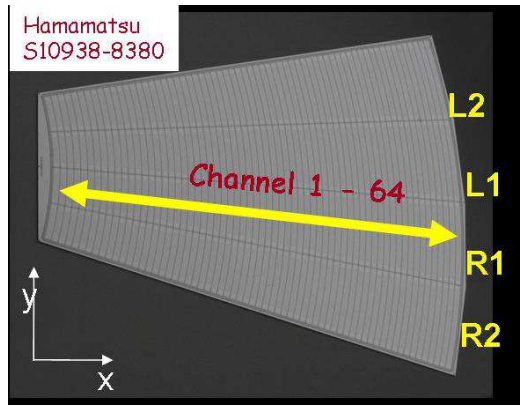}
  \caption{A prototype silicon sensor for LumiCal}
  \label{fig:sensor_lumi}
\end{figure}

\begin{figure}[htpb]
  \centering 
  \subfigure[]{
  \label{fig:mb2a} 
      \includegraphics[width=0.45\textwidth]{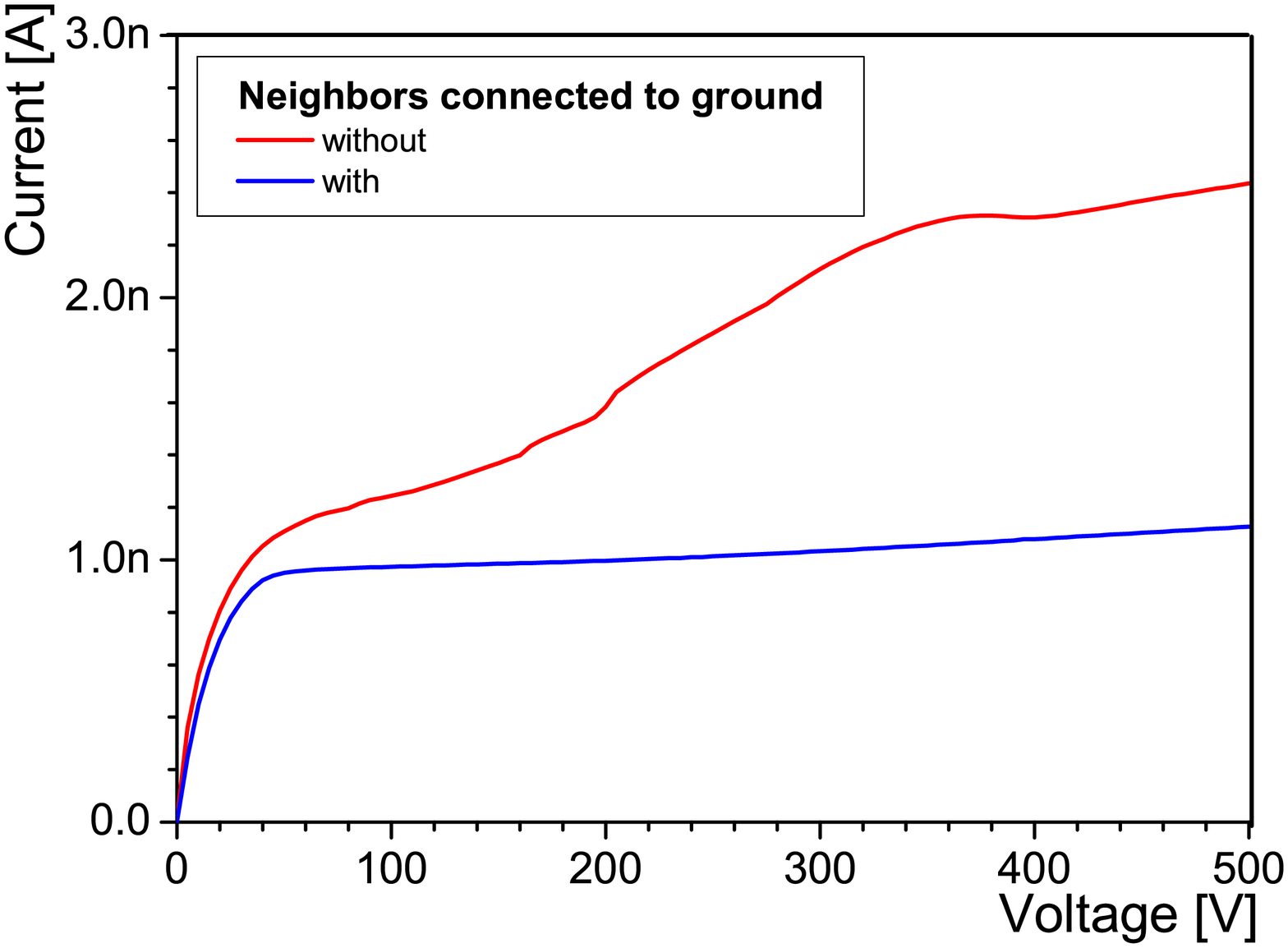}} 
  \subfigure[]{
     \label{fig:mb2b}
      \includegraphics[width=0.45\textwidth]{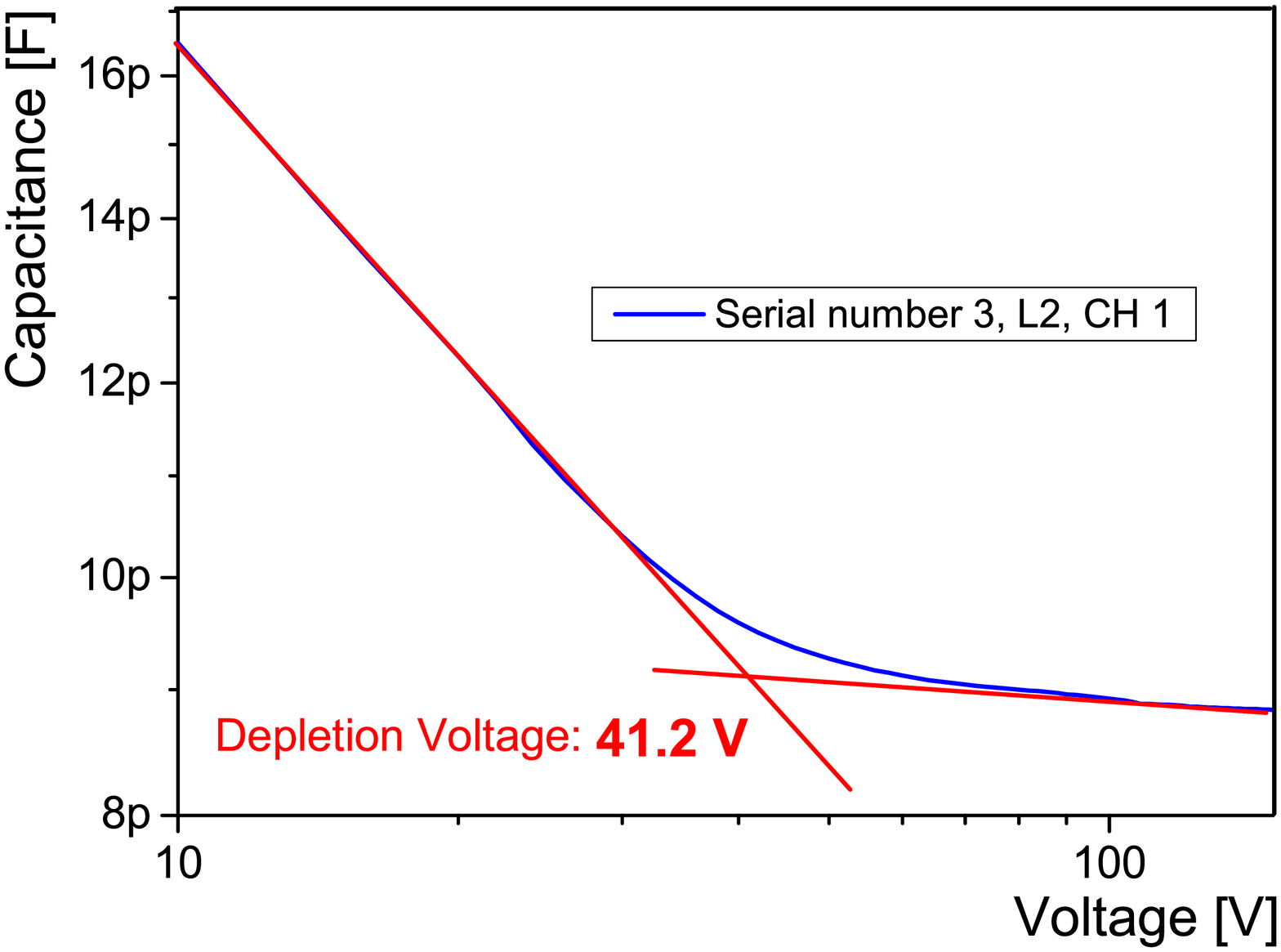}}
   \caption{
\Subref{fig:mb2a} The dependence of the leakage current on the bias voltage 
for a single pad with and without grounded
  neighbours.
\Subref{fig:mb2b} The capacitance of a pad as a function of the bias voltage.}
  \end{figure}

Prototypes of LumiCal sensors have been designed~\cite{eudet_sensors} 
and then manufactured by Hamamatsu Photonics. A picture 
of a sensor is shown in Figure~\ref{fig:sensor_lumi}. Its shape is a ring segment
of 30$^\circ$. The thickness of the n-type silicon bulk is 320 $\mu$m. The pitch 
of the concentric 
p$^+$ pads is 1.8 mm and the gap between two pads is 0.1 mm.
The leakage current of a single pad
as a function of the bias voltage
is shown in Figure~\ref{fig:mb2a}. 
Putting the neighbouring pads on ground stabilises the measurement
and reduces the current values by a factor of two. 
The leakage currents  
of all the pads of one sensor
have been measured at a bias voltage of 500~V. 
All pads except one have a leakage
current in the range from 1 to 4~nA. Less than 5\% of all pads have 
a break-through voltage below 500~V. For other 
sensors the results are similar.
The capacitance as a function of the bias voltage for a pad is shown in Figure
\ref{fig:mb2b}. Also shown is how the value of the full
depletion voltage is obtained. Values from 39~V to 43~V were found. 
At a voltage of 100~V the pad capacitance values are
between 8~pF for the smallest pads and 25~pF for the largest pads.

%% file: asic_lumi_jinst.tex
\section{ASIC developments}

Since the occupancy in BeamCal and LumiCal is relatively large
they must be read-out after each bunch crossing. Therefore special
front-end and ADC ASICs have been developed which match the timing 
of the ILC -- bunch trains with a frequency of 5 Hz
and about one ms duration with 300 ns between bunches.
Since the ASICs are positioned at the outer radius of the calorimeters 
the expected radiation dose is noncritical.
From Monte Carlo simulations less than 140 Gy and about one Gy are estimated 
for BeamCal and LumiCal, respectively, for one year of operation at 
500 GeV centre-of-mass energy and nominal beam parameters.      

\subsection{LumiCal readout}

The design of the LumiCal front-end electronics 
was performed
for the proposed detector architecture~\cite{VFCAL_memo}. 
The front-end ASIC 
is supposed to
work in two modes, the physics mode and the calibration mode. 
In the physics mode, electromagnetic showers will be measured
with large energy depositions on the pads. The front-end ASIC 
must
process signals 
up to at least 6~pC per channel.  
In the calibration mode, MIP signals from single relativistic muons
will
be measured. 
The minimum size of these signals is 2~fC, corresponding to the low end of the Landau 
distribution for MIPs in 300~$\mu$m thick silicon.
From the sensor segmentation a range of pad 
capacitances between 10~pF and 100~pF was obtained\footnote{The sensor segmentation was revised 
later, resulting in pad capacitances between 10~pF and 25 pF.}. 
Because of the high expected occupancy, the front-end ASIC 
needs to
be fast enough to resolve signals from subsequent bunch crossings which 
are separated in time by about 300~ns.

The simulations of LumiCal indicate that the shower reconstruction needs 
at least 8~bit precision.
Severe requirements set on the readout electronics power dissipation
may be strongly relaxed if switching of the power between bunch trains 
is done.
This is feasible since in the ILC experiments after
each 1~ms bunch train there will be
a pause of about 200~ms~\cite{nominal_set}.

\begin{figure}[htb!]
\centering 
  \subfigure[]{\label{fig:FEfoto} 
   \includegraphics[width=0.45\columnwidth]{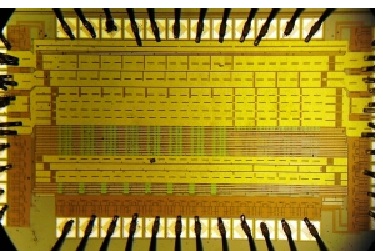}}
 \subfigure[]{\label{fig:ADCfoto}
\includegraphics[width=0.45\columnwidth,height=0.3\columnwidth]{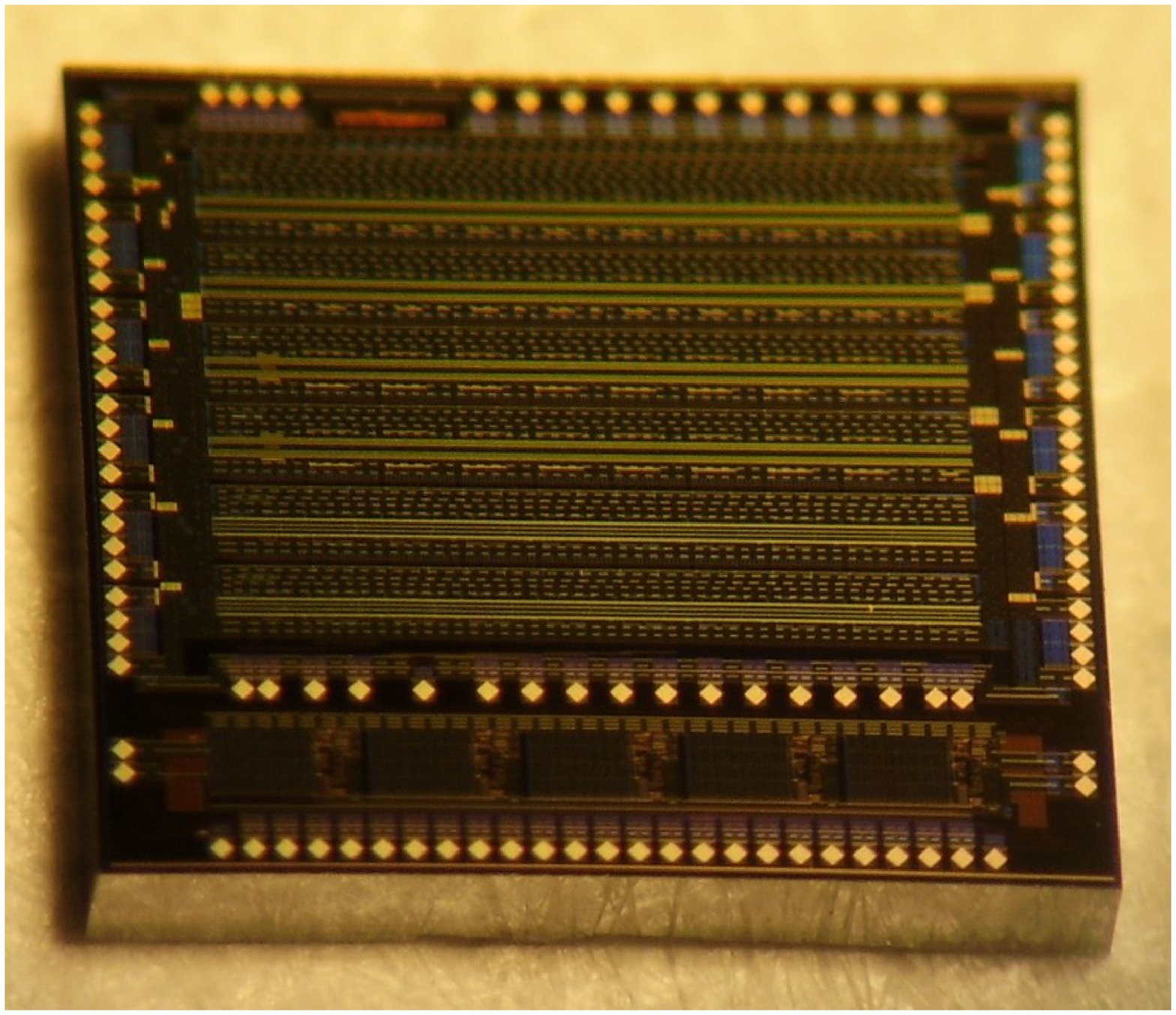}}
\caption{
    Photograph of prototypes of the front-end ASIC \Subref{fig:FEfoto} and 
     the ADC ASIC \Subref{fig:ADCfoto}}
\end{figure}
The prototype ASICs,
as shown in Figures~\ref{fig:FEfoto} and~\ref{fig:ADCfoto}, are fabricated in 0.35~$\mu$m CMOS technology. 

\subsubsection{Front-end electronics design}
\begin{figure}[htb!]
\centering 
\centerline{\includegraphics[width=0.7\columnwidth]{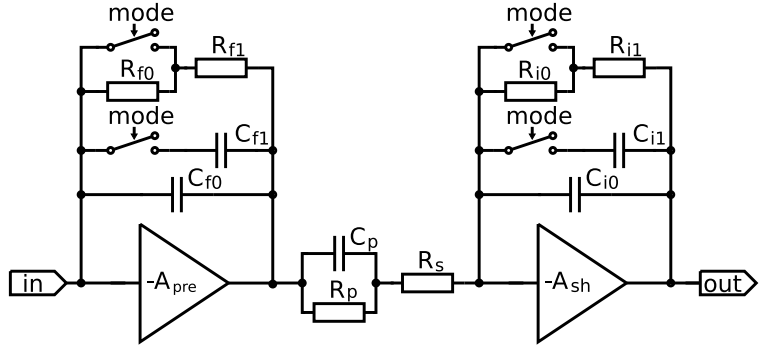}}
\caption{Block diagram of the single front-end channel}\label{fig:frontend}
\end{figure}

The chosen front-end architecture comprises a charge sensitive amplifier, 
a pole-zero cancellation circuit (PZC) and a shaper, as shown in 
Figure~\ref{fig:frontend}.
In order to cope with large charges in the physics mode and small ones 
in the calibration mode a variable gain in both the charge amplifier and 
the shaper is applied. The mode switch in Figure~\ref{fig:frontend} changes 
the effective values of the feedback circuit components 
${\rm{R}}_{\rm{f}}$, ${\rm{C}}_{\rm{f}}$, ${\rm{R}}_{\rm{i}}$, ${\rm{C}}_{\rm{i}}$ 
and 
therefore 
the transimpedance gain of the front-end ASIC
is changed.
The low gain (large ${\rm{C}}_{\rm{f}}$) is used for the physics mode when the front-end
processes signals from large charge depositions in the sensor, while the high
gain (small ${\rm{C}}_{\rm{f}}$) is used in the calibration mode.
Assuming high enough open loop gain of the pre-amplifier (${\rm{A}}_{\rm{pre}}$) 
and the shaper amplifier (${\rm{A}}_{\rm{sh}}$), the transfer function of this circuit 
is given 
by
\begin{equation}
\begin{array}{l}
\displaystyle \frac{{\rm{U}}_{\rm{out}}(s)}{{\rm{I}}_{\rm{in}}(s)}  = \frac{1}{{\rm{C}}_{\rm{f}} {\rm{C}}_{\rm{i}} {\rm{R}}_{\rm{s}}} \cdot \frac{s+1/{{\rm{C}}_{\rm{p}} {\rm{R}}_{\rm{p}}}}{s+1/{{\rm{C}}_{\rm{f}} {\rm{R}}_{\rm{f}}}} \cdot 
\displaystyle \frac{1}{(s+1/{{\rm{C}}_{\rm{i}} {\rm{R}}_{\rm{i}}}) (s+1/{{\rm{C}}_{\rm{p}} ({\rm{R}}_{\rm{p}} || {\rm{R}}_{\rm{s}})})}.
\end{array}
\label{eq:us1}
\end{equation}
By setting properly
the PZC parameters
(${\rm{C}}_{\rm{f}} {\rm{R}}_{\rm{f}} = {\rm{C}}_{\rm{p}} {\rm{R}}_{\rm{p}}$) and 
by
equalising 
the
shaping time constants (${\rm{C}}_{\rm{i}} {\rm{R}}_{\rm{i}} = {\rm{C}}_{\rm{p}} ({\rm{R}}_{\rm{p}} || {\rm{R}}_{\rm{s}})$), 
one obtains the first 
order shaping, equivalent to a CR-RC filter, with a peaking time ${\rm{T}}_{\rm{peak}}={\rm{C}}_{\rm{i}}{\rm{R}}_{\rm{i}}$.
A simple first order shaping is chosen as a trade-off between the noise and
the power dissipation. 
Regarding the noise,
the
main requirement is to obtain in calibration
mode the signal to noise ratio of about 10 for the largest sensor capacitances.
Both of the amplifying stages (${\rm{A}}_{\rm{pre}},{\rm{A}}_{\rm{sh}}$) are designed as folded 
cascodes~\cite{amplex} with active loads, followed by source followers. 
In the prototype ASIC,
eight 
front-end channels are implemented.
A more detailed discussions of the front-end ASICs can be found in Ref.~\cite{fe_nim}.

\subsubsection{Front-end electronics measurements}
\begin{figure}[htb!]
\centering 
  \subfigure[]{\label{fig:pulses} 
   \includegraphics[width=0.35\columnwidth,height=0.45\columnwidth,angle=-90]{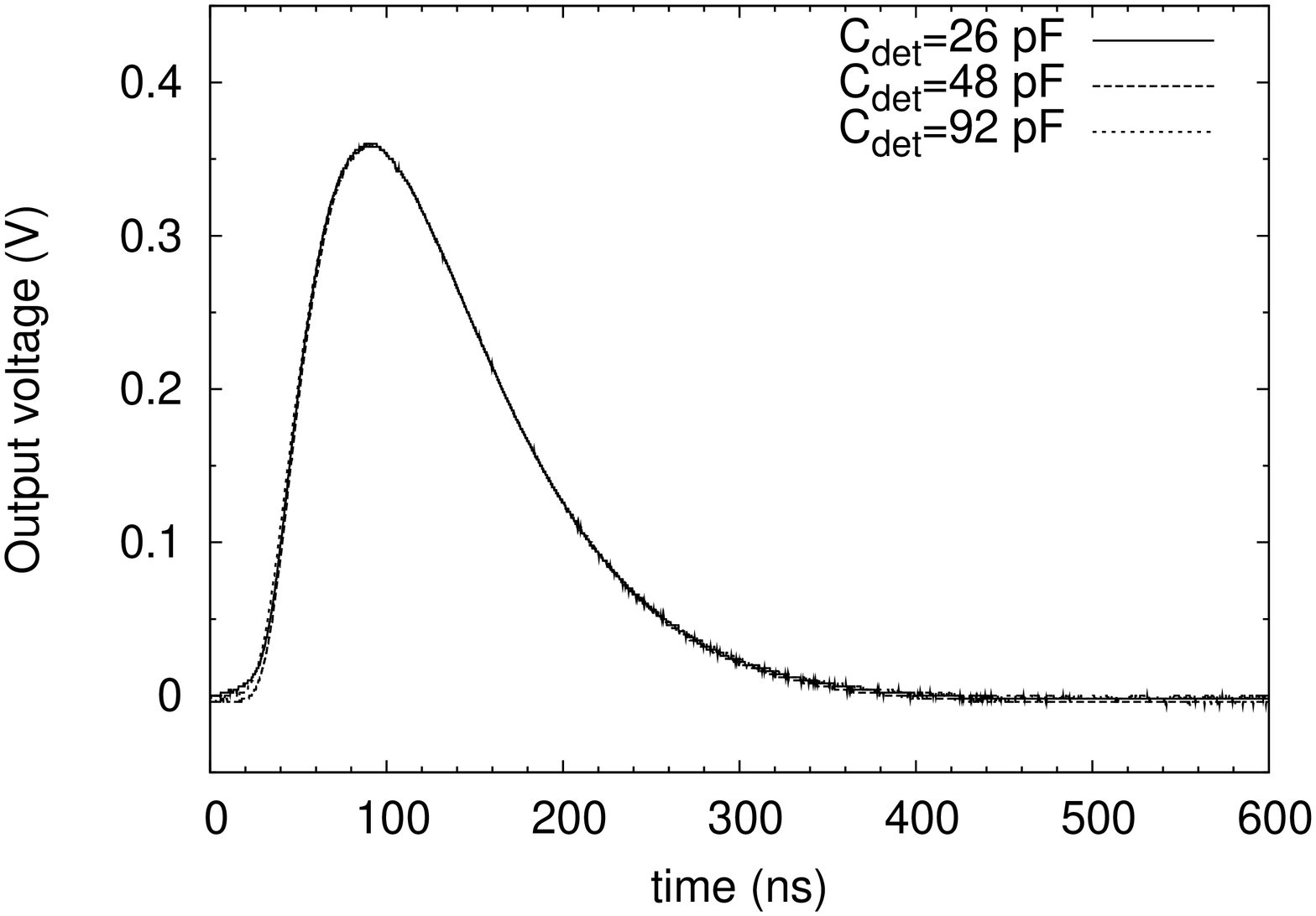}}
  \subfigure[]{\label{fig:noise_rms}
   \includegraphics[width=0.38\columnwidth,height=0.5\columnwidth,angle=-90]{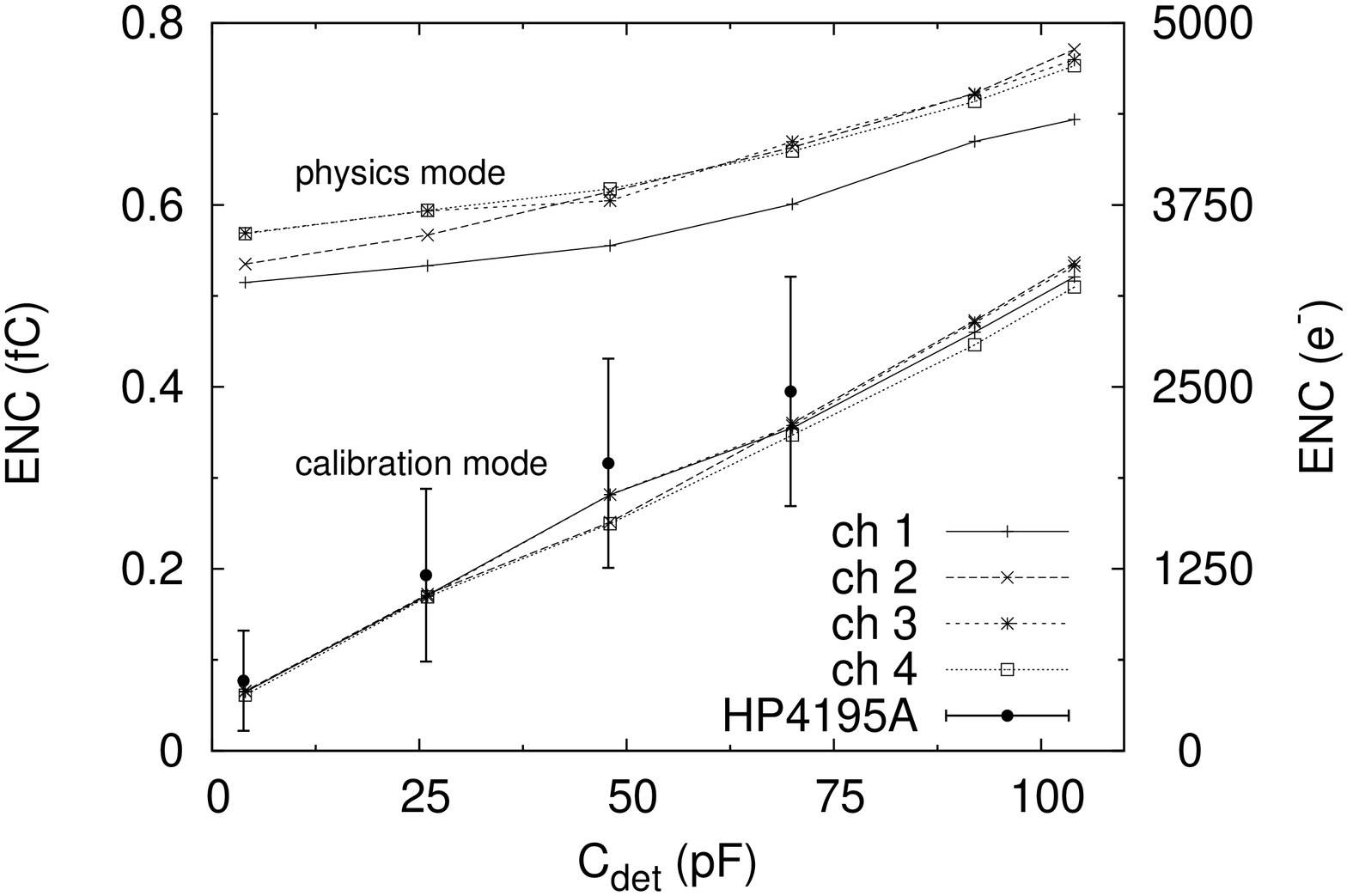}}
\caption{
\Subref{fig:pulses}
Output pulses in physics mode 
as a function of the input capacitance
for ${\rm{Q}}_{\rm{in}}$=3.3~pC. 
\Subref{fig:noise_rms}
Noise ENC measurements obtained with true r.m.s. meter for the front-end ASIC.}
\end{figure}

Figure~\ref{fig:pulses} shows the response of the front-end channel
to charge injected through the input test capacitance for different 
values of the input capacitance, ${\rm{C}}_{\rm{det}}$, within the interesting range. 
The sensor capacitance is simulated with an external capacitor.
It is seen that both the amplitude and the peaking 
time are not sensitive to the value of the input capacitance
in agreement
with 
HSPICE
simulations.

The output noise has been measured using a HP3400 true r.m.s. meter~\cite{agilent}.
The equivalent noise charge, ENC, as a function of input capacitance is shown 
in Figure~\ref{fig:noise_rms}. 
Results obtained for the physics and calibration modes are 
shown on the same plot.
Since the HP3400 bandwidth is only up to 10~MHz the numbers may be underestimated
by about 20\%. The measured ENC as a function of ${\rm{C}}_{\rm{det}}$ are
in agreement with simulations. In particular, in the calibration mode the signal 
to noise ratio of 10 is maintained for input capacitances up to about 100~pF. 
For a few points additional noise measurements have been
performed by measuring the output noise spectra using a HP4195A spectrum analyser~\cite{agilent}
and then integrating it numerically.
The results of such measurements are added in Figure~\ref{fig:noise_rms}.
They agree within their uncertainties with the HP3400 measurements.

In order to test the effectiveness of the PZC circuit, the front-end response has been
measured as a function of the rate of input pulses.
To avoid input charges of both polarities when using a square-wave test signal,
the staircase test waveforms are synthesised using the Tektronix AWG2021 waveform generator.
It was found that the change in amplitude reaches 2\%
for input rates of about 3~MHz and is 
quite insensitive
to
the input capacitance. 
The power consumption of about 8.9~mW/channel is measured 
in accordance with expectations from simulation.

\subsubsection{ADC design}
As a compromise
between speed, area and power consumption the ADC was designed using pipeline
technology.
A 1.5-bit per stage architecture is chosen because of
its simplicity and immunity to the offsets in the comparator and amplifier circuits.
The prototype ADC consists of an input sample and hold
circuit, 9 pipeline stages and digital correction circuitry. In addition, the power
switching feature is also implemented.
More details about the ADC design can be found in Ref.~\cite{adc_mixdes}.

\subsubsection{ADC performance measurements}

\begin{figure}[htb!]
\centering
  \subfigure[]{\label{fig:inl_static} 
   \includegraphics[width=0.49\columnwidth]{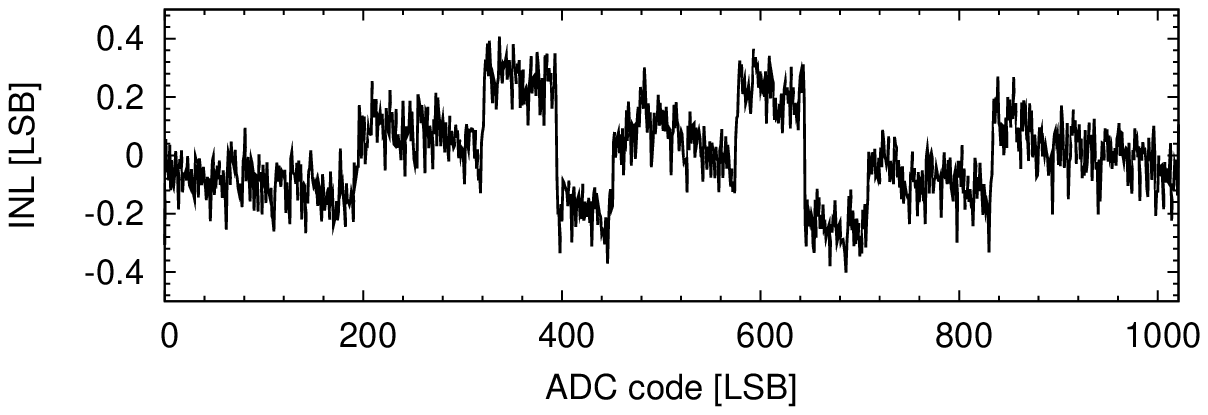}}
  \subfigure[]{\label{fig:dnl_static} 
   \includegraphics[width=0.49\columnwidth]{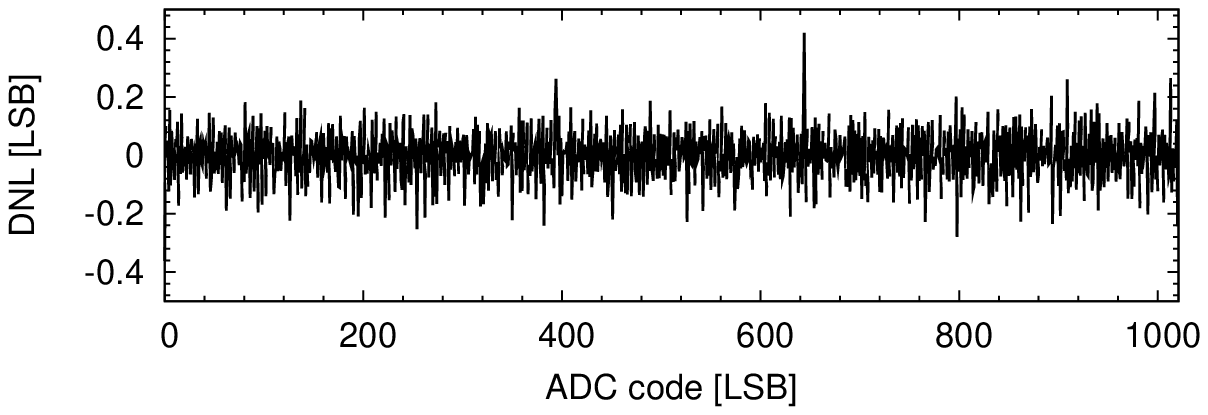}}
\caption{Static measurements of \Subref{fig:inl_static} INL and 
        \Subref{fig:dnl_static} DNL 
        at 20 MHz sampling frequency.}
\end{figure}

\begin{figure}[htb!]
\centering
  \subfigure[]{\label{fig:dyn_fft} 
   \includegraphics[width=0.3\columnwidth,height=0.49\columnwidth,angle=-90]{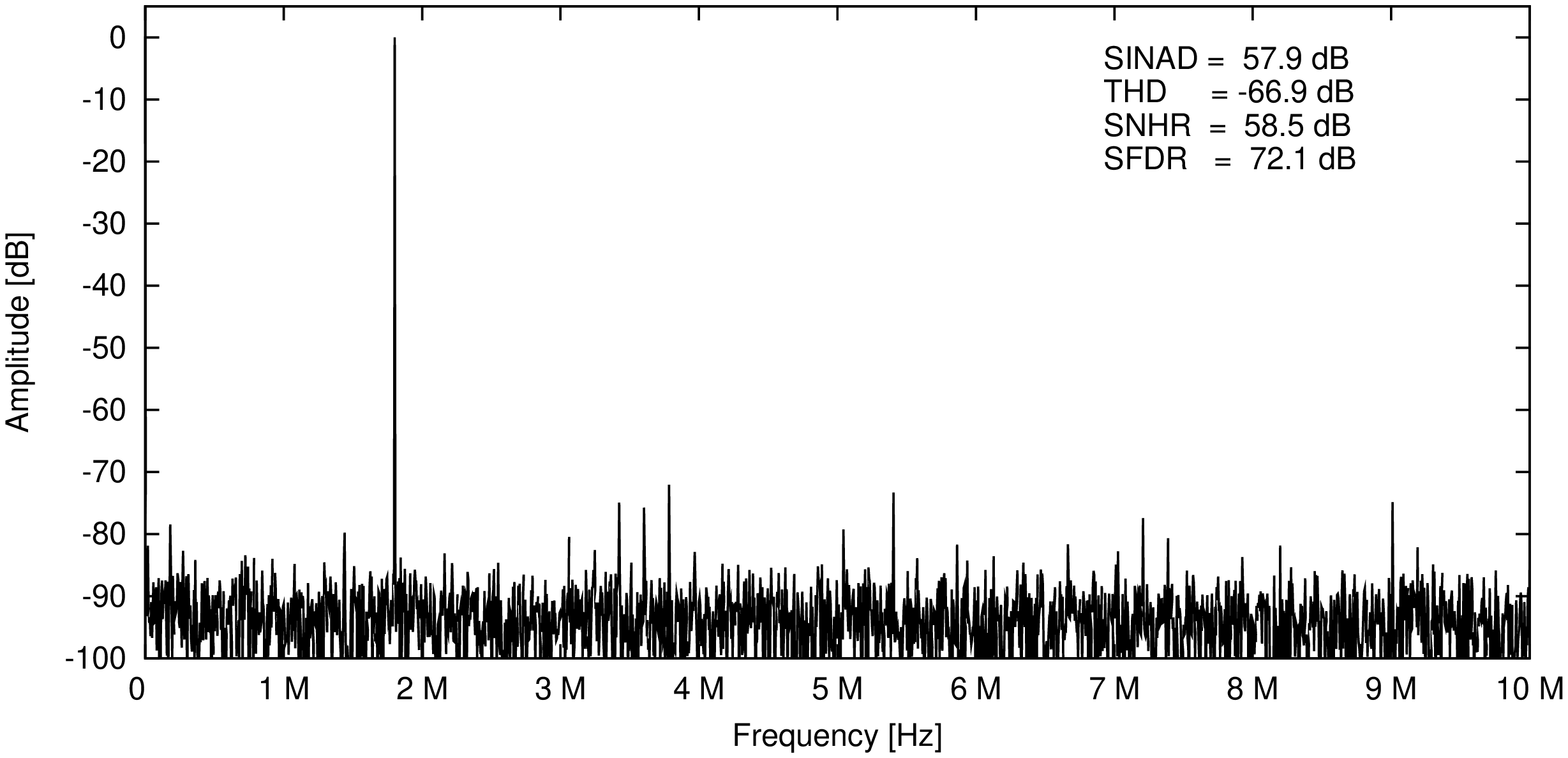}}
  \subfigure[]{\label{fig:dyn_fadc} 
    \includegraphics[width=0.3\columnwidth,height=0.49\columnwidth,angle=-90]{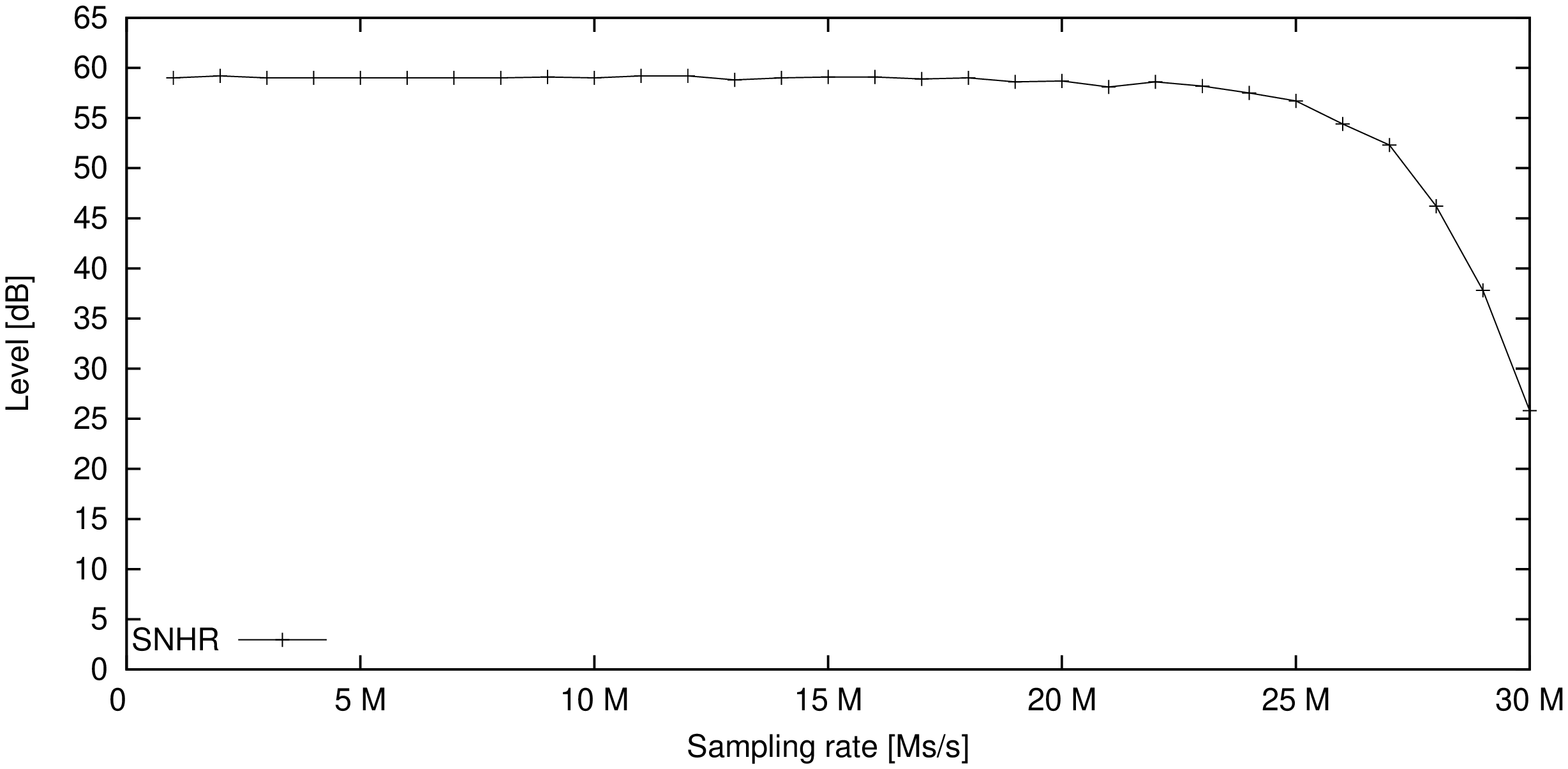}}
   \caption{\Subref{fig:dyn_fft} Example of the Fourier spectrum
           measurement with ${\rm{f}}_{\rm{in}}$=1.8~MHz 
           and  ${\rm{f}}_{\rm{clk}}$=20~MHz,
         \Subref{fig:dyn_fadc}
         SNHR as a function of the sampling rate.}
\end{figure}

The static measurements of the Integral Nonlinearity, INL, and the Differential Nonlinearity, DNL, 
obtained at a sampling frequency of 20~MHz, are shown in Figures~\ref{fig:inl_static}
and~\ref{fig:dnl_static},
respectively.
These parameters are calculated using the histogramming method.
The measured INL is always less than 1~LSB while the
DNL is below 0.5~LSB. 
These results attest to 
a
very good ADC linearity.
To estimate the dynamic performance, measurements with sinusoidal wave 
input are performed \cite{ieee}. 
An example of a measured Fourier spectrum using a 1.8~MHz full scale (0~dB) input signal sampled at 20~MHz 
is shown in Figure~\ref{fig:dyn_fft}. 
It is seen that the noise and harmonic components are small enough
not to
affect significantly
the resolution.
The signal to noise ratio, SNHR, is measured as a function 
of sampling frequency as shown in Figure~\ref{fig:dyn_fadc}. 
An SNHR of about 58~dB is obtained in the frequency range up to almost 25~MHz.

%% file: BeamCal_ASIC.tex
\subsection{BeamCal readout}

The BeamCal ASIC, designed for 180 nm TSMC technology, will be able to handle 
32 channels. The two modes of operation require a front-end circuit capable of a wide performance 
envelope: high slew rate for standard data taking, and low noise for calibration. 
In standard data taking the occupancy is high, and therefore all data from a full bunch train 
must be recorded, to be read out between bunch trains. Because of its reliability, density 
and redundancy, 
a digital memory array will be used to store the data from 
all collisions in each bunch train. This choice requires a sampling rate of 3.25 MHz per 
channel, which is achieved by 10-bits, successive approximation analog-to-digital 
converters \cite{McCreary100}. 
The small size of this ADC architecture allows to use one converter per channel.

In this front-end ASIC, the dominant noise source is the charge sensitive amplifier 
series noise. Assuming 40 pF input 
capacitance, high occupancy and the 300 ns period,
a careful design 
of noise filtering and baseline restoration
is
necessary\cite{Spieler500}.

In order to take advantage of all the time available for signal processing, the filter 
for calibration operation has been implemented using switched-capacitor, SC, 
circuits \cite{Gregorian100}. This technique allows to precisely define the 
circuit time constants depending on the input clock frequency and the ratio 
of two capacitors. Baseline restoration is achieved by means of a fast gated reset, 
followed by a slow reset-release technique to reduce the effect of a split doublet. 
The slow reset-release is implemented using SC circuits.

In standard data taking operation, an adequate noise power is effectively achieved 
by means of a slow reset-release technique, similar to that used in calibration 
operation. An explicit filter for standard data taking operation is unnecessary, 
as the amplifier bandwidth suffices for noise filtering purposes.

\begin{figure}
	\centering
		\includegraphics[width=0.7\textwidth]{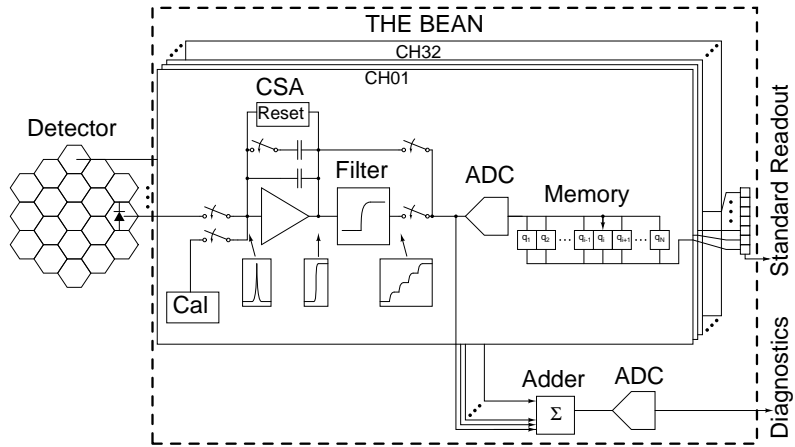}
	\caption{Simplified BeamCal ASIC block diagram of a single channel. In addition to the
        standard read-out a fast analog sum of groups of pads for beam-tuning is delivered by
        the Adder.}
	\label{fig:Updated_Block_Diagram}
\end{figure}

Figure \ref{fig:Updated_Block_Diagram} shows a simplified block diagram for a 
single channel. In standard data taking operation, 
since  filtering is unnecessary, 
the integrator is bypassed to reduce power consumption.

For design purposes, the transistor-level noise analysis has been carried out 
using the ${\rm{g}}_{\rm{m}}/{\rm{I}}_{\rm{D}}$ technique \cite{Silveira100}, 
which takes noise coefficients 
directly from SPICE simulation results. As this is a gated front-end, the system-level 
noise analysis has been done using the weighting function approach.

\begin{figure}
	\centering
		\includegraphics[width=0.90\textwidth]{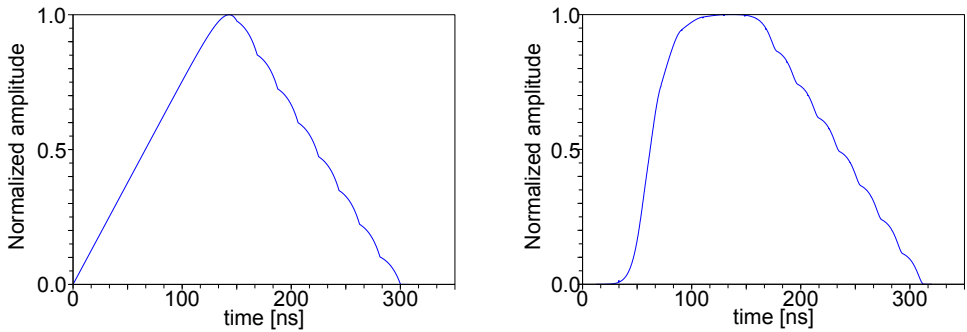}
	\caption{Front-end weighting function assuming ideal 
                 components (left) and simulation results (right) in the calibration mode}
	\label{fig:Combined_W_Fcns_Small}
\end{figure}

Since the system's dominant noise source is series noise, a triangular-shaped weighting 
function effectively minimises the output noise power. The negative slope section of 
the triangular weighting function is easily implemented by means of an integrator -- in 
this case, a SC integrator. The positive slope section is achieved by means of the slow 
reset-release technique mentioned earlier. The weighting function resulting from an ideal 
reset-release and a SC integrator is shown in Figure \ref{fig:Combined_W_Fcns_Small}, 
left; a more realistic weighting function, reconstructed from SPICE simulation 
results, is shown in the right plot. In both cases, the target noise level is effectively achieved.

\subsubsection{Circuit implementation}

The charge sensitive amplifier is a folded-cascode amplifier with NMOS input device, connected to a 
switched-capacitor feedback network. The amplifier input transistor is biased at 
450 {$\mu$A} 
whereas the load works at about 
50 $\mu$A. The feedback network consist of two feedback 
capacitors of 
0.9 pF
and 
44.1 pF
for calibration and standard data taking modes, 
respectively. Both have a reset transistor, with a gate voltage driven by the switched-capacitor 
reset-release network. The amplifier output is pseudo-differential.

In order to isolate the amplifier from the filter's SC-related kickback noise, a buffer 
circuit is used. The buffer also allows signal shifting, producing a more adequate 
common-mode level for the filter. The buffer consumes 
130 $\mu$A and
consists of a source follower, with cascoded current source and an additional device 
to keep a nearly constant operational point in the input transistor. This serves the purpose of 
enhancing the buffer linearity.

The filter implemented is a fully-differential switched-capacitor integrator. 
Capacitor values were carefully designed in order to obtain the adequate noise 
performance. The core of the integrator is a class A/AB amplifier \cite{Rabii500} 
that consumes 
456 $\mu$A.

The converter is a 10-bit, fully-differential successive approximation register ADC. 
The one included in the BeamCal ASIC has 16 fF unit capacitances, and similar 
versions with 4 fF and 2 fF unit capacitances were also designed for individual characterisation.

\begin{figure}
	\centering
		\includegraphics[width=0.30\textwidth]{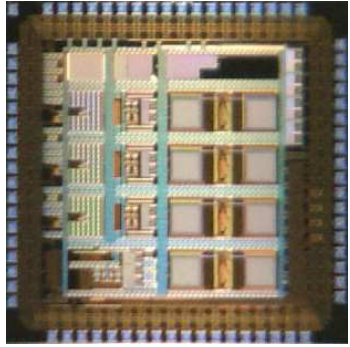}
	\caption{BeamCal Instrumentation ASIC Prototype}
	\label{fig:BeanPic}
\end{figure}

The BeamCal ASIC prototype, similar to the ASIC described in 
Figure \ref{fig:Updated_Block_Diagram}, but including only three channels and no 
internal memory, was fabricated and is currently being tested. Figure \ref{fig:BeanPic} 
shows the 
2.4 mm $\times$ 2.4 mm die. 

\subsubsection{Test results}

\begin{figure}[htb!]
\centering
   \subfigure[]{\label{fig:inl_beamcal}        
		\includegraphics[width=0.49\textwidth]{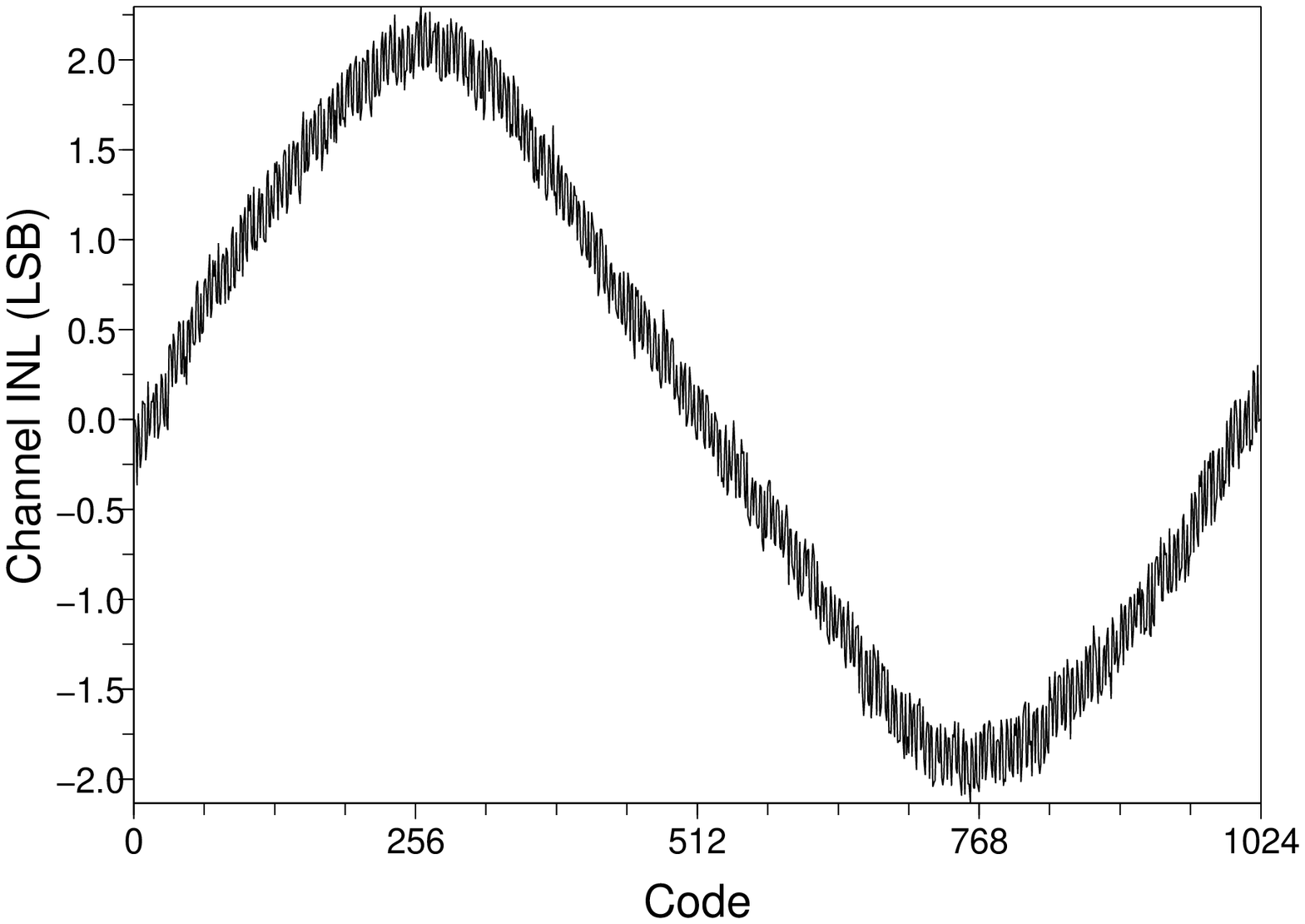}}
   \subfigure[]{\label{fig:dnl_beamcal}
		\includegraphics[width=0.49\textwidth]{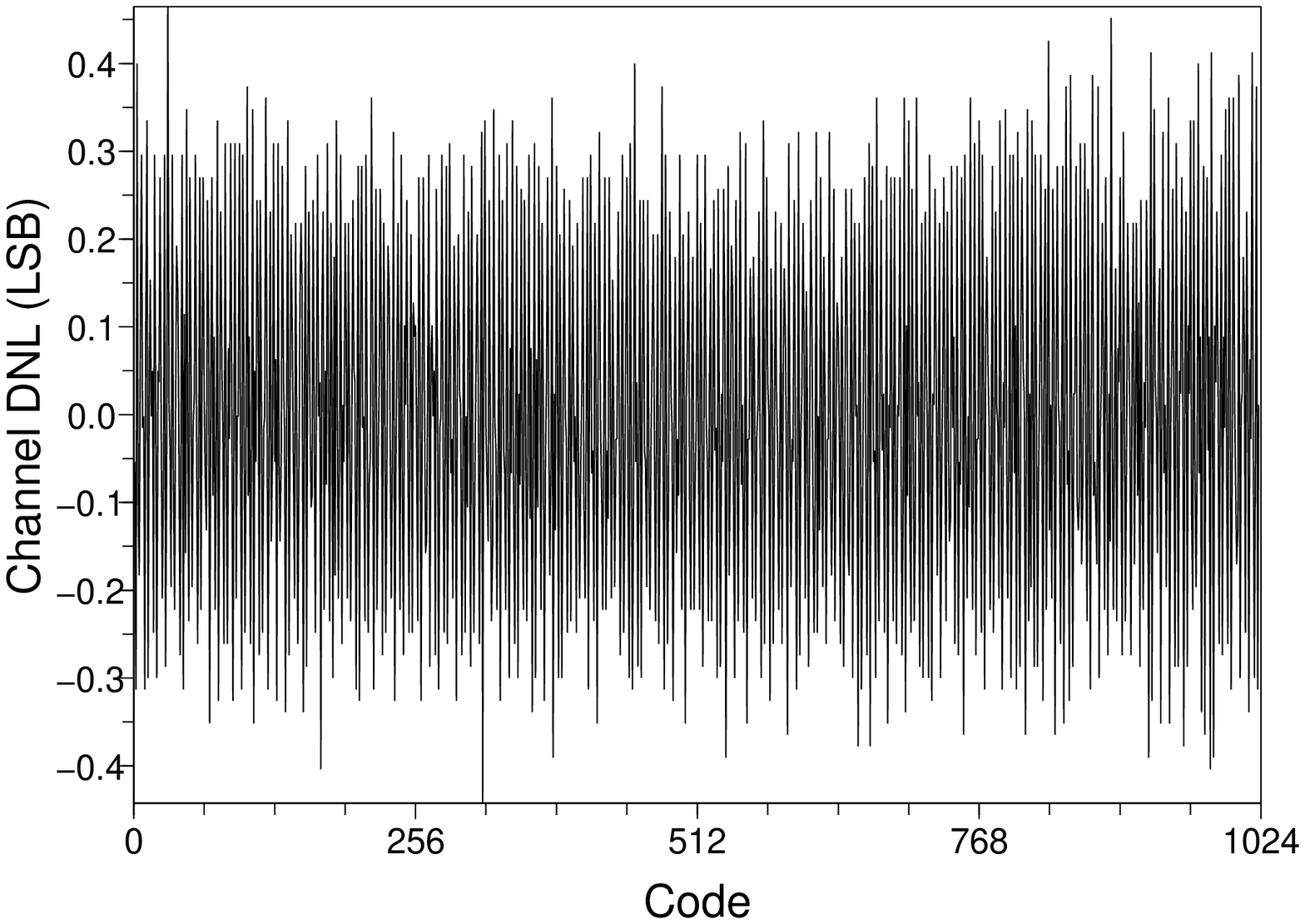}}
	\caption{Results of \Subref{fig:inl_beamcal} the INL and 
        \Subref{fig:dnl_beamcal} the DNL
        using 2 fF unit capacitors.}
\end{figure}

The ADC in the BeamCal ASIC has been quantitatively characterised, 
along with the additional versions of the ADC using smaller unit capacitances.
Figures~\ref{fig:inl_beamcal} and~\ref{fig:dnl_beamcal} show the 
INL and DNL for the ADC using 2 fF capacitors. The measurements were done at the nominal 
sampling frequency of
3.125MHz. The ADC input was a ramp, generated by 16-bit DAC, and the static performance 
measurements were calculated using the histogram method on
the ADC digital output. The results are consistent 
with unit capacitance matching better than 0.1\%. 
The INL cubic-like shape in Figure~\ref{fig:inl_beamcal} is
explained due to copper dishing effects, and will be corrected in 
future versions by re-arranging the capacitor array connections.

%% file: pair_moi_asic.tex
\subsection{Pair monitor readout}

\begin{figure}[hbt]
\begin{center}
\includegraphics*[width=11cm]{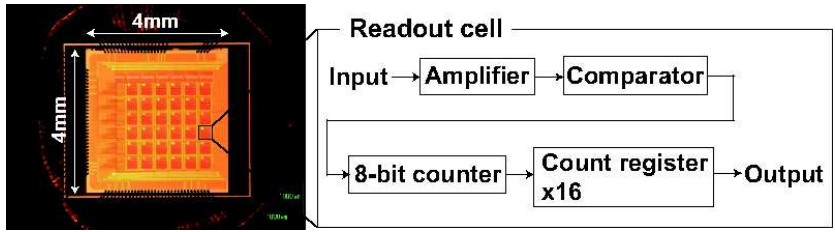}
\end{center}
\caption{\label{fig:chip}
Picture of the prototype of the pair monitor 
readout ASIC and schematic diagram of the circuit in a readout cell. 
The readout cell consists of the amplifier, 
comparator, 8-bit counter, and 16 count-registers.
}
\end{figure}

A prototype ASIC has been designed
with 36 readout cells arranged as an array 
of 6$\times$6, as shown in Figure \ref{fig:chip}. Each cell has an amplifier block, 
comparator, an 8-bit counter and a 16 count-registers. The amplifier block consists 
of a charge sensitive pre-amplifier, a threshold block and a differential-amplifier. 
The pre-amplifier is a constant-current feedback-type amplifier. The time-over-threshold 
of the output signal is proportional to the injected charge through the constant 
current feedback in the pre-amplifier. In the 8-bit counter, the Gray code is used to 
count the number of hits. The 16 count-registers are prepared to store hit counts
in one bunch train subdivided in 16 time slices. 
There are also decoders which select a count-register 
to store and readout the hit count. A shift register to select a readout pixel, data 
transfer to the output line and distributor of the operation signals are arranged 
around the 36 readout cells as a glue logic. The bonding pad is prepared in each 
cell to be attached to a sensor with bump bonding. The prototype ASIC has been 
produced with TSMC 250 nm CMOS process. The chip size is 4$\times$4 mm$^{2}$, 
and the readout cell size is 400$\times$400 $\mu$m$^{2}$. 

\begin{figure}[hbt]
\begin{center}
\includegraphics*[width=10cm]{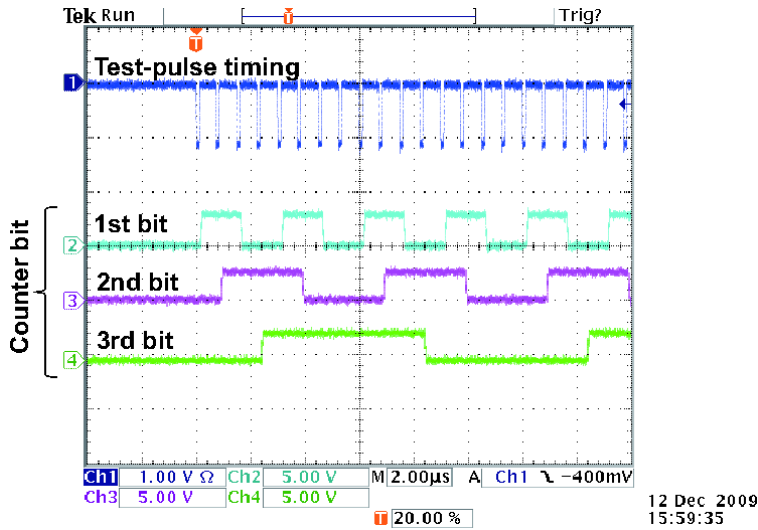}
\end{center}
\caption{\label{fig:counter}
Output signals from the counter block.
The lower 3 bits of the 8-bit counter are shown. The test-pulse timing corresponds to the bunch crossing frequency if the
ILC.
}
\end{figure}

Figure \ref{fig:counter} shows the response of the counter block. The state of the counter bits changes
at each test pulse indicating a bunch crossing. 
The number of hits
is measured in 16 time slices of a bunch train.
The data stored will then 
be
read-out during the inter-train time.
The test is performed counting the hits in each time slice with 
a count rate of 4 MHz, larger than expected at the ILC.
The number of hits was counted without any bit lost.

We also studied the noise level in the circuit. The count efficiency was investigated as a 
function of the threshold voltage at the comparator. 
Fitting the efficiency curve with the error function, 
a standard deviation of 0.94 mV was obtained. 
With the gain of $1.6 \times 10^{-3}$ mV per electron, this corresponds 
to an ENC of about 600 electrons.

As the next step, a pair-monitor prototype will 
be
built 
in Silicon On Insulator technology. 
The sensor and readout ASIC will be prepared on 
the same wafer. This prototype will be used
to investigate 
not only the standard characteristics but also the radiation tolerance.
Currently, an ASIC is developed in OKI 0.2 $\mu$m FD-SOI 
CMOS~\cite{oki_fd-soi} technology.